\documentclass[sigconf]{acmart}
\renewcommand\footnotetextcopyrightpermission[1]{}
\usepackage{booktabs} 
\usepackage[ruled,linesnumbered]{algorithm2e}
\usepackage{booktabs}
\usepackage[british]{babel}
\usepackage{paralist}
\usepackage{multirow}
\usepackage{amsmath}
\usepackage{amssymb}
\usepackage[marginal]{footmisc}
\usepackage{subfigure}
\usepackage{graphicx}
\usepackage{balance}
\usepackage{booktabs}
\usepackage{threeparttable}
\usepackage{hyperref}
\usepackage{color}
\usepackage{colortbl}
\usepackage{array}
\usepackage{comment}
\usepackage{float}
\usepackage{url}
\usepackage{xcolor}
\usepackage{alltt}
\usepackage{framed}
\usepackage{enumitem}

\SetCommentSty{mycommfont}
\usepackage{algpseudocode}
\usepackage{tikz}
\usetikzlibrary{trees}
\usepackage{breqn}
\usepackage{pgfplots}
\usepackage{bchart}
\usepackage{caption} 
\newcommand*{\affmark}[1][*]{\textsuperscript{#1}}

\newcommand{\ifdepth}{\emph{if-depth}}
\newcommand{\nestif}{\emph{nested-IF}}
\newcommand{\nestifformu}{\emph{nested-IF} formulae}
\newcommand{\nestiffunc}{\emph{nested-IF} functions}
\newcommand{\nestifexp}{\emph{nested-IF} expressions}

\setcopyright{rightsretained}

\makeatletter
\def\@copyrightspace{\relax}
\makeatother

\begin{document}
\title{Automated Refactoring of Nested-IF Formulae in Spreadsheets}
\author{Jie Zhang\affmark[1] , Shi Han\affmark[2], Dan Hao\affmark[1], Lu Zhang\affmark[1], Dongmei Zhang\affmark[2]}
\affiliation{%
  \institution{\affmark[1]Key Laboratory of High Confidence Software Technologies (Peking University), MoE, Beijing, China\\\affmark[2]Microsoft Research, Beijing, China}
\email{\affmark[1]\{{zhangjie_marina,haodan,zhanglucs\}}@pku.edu.cn} 
\email{\affmark[1]\{zhangjie\_marina,haodan,zhanglucs\}@pku.edu.cn,
\affmark[2]\{shihan,dongmeiz\}@microsoft.com}
}

%
%
%
%
%


\begin{abstract}
Spreadsheets are the most popular end-user programming software, where formulae act like programs and also have smells. One well recognized common smell of spreadsheet formulae is \nestifexp{}, which have low readability and high cognitive cost for users, and are error-prone during reuse or maintenance. However, end users usually lack essential programming language knowledge and skills to tackle or even realize the problem. The previous research work has made very initial attempts in this aspect, while no effective and automated approach is currently available. 

This paper firstly proposes an AST-based automated approach to systematically refactoring \nestifformu{}. The general idea is two-fold. First, we detect and remove logic redundancy on the AST. Second, we identify higher-level semantics that have been fragmented and scattered, and reassemble the syntax using concise built-in functions. A comprehensive evaluation has been conducted against a real-world spreadsheet corpus, which is collected in a leading IT company for research purpose. The results with over 68,000 spreadsheets with 27 million \nestifformu{} reveal that our approach is able to relieve the smell of over 99\% of \nestifformu{}. Over 50\% of the refactorings have reduced nesting levels of the \nestif{}s by more than a half. In addition, a survey involving 49 participants indicates that for most cases the participants prefer the refactored formulae, and agree on that such automated refactoring approach is necessary and helpful.
\end{abstract}

%
%
%


\maketitle

\section{Introduction}
\label{sec:introduction}

Spreadsheets are the most popular end-user programming tools~\cite{spreadmostpopular}. One of the most important enabling factors is that spreadsheets provide immediate feedback so users can make a change in one place and immediately see the results~\cite{burnett10}. Underneath such an advantage, formulae play an important role as end-user friendly programs. However, end-users typically lack essential knowledge and skills of programming, and are easier to write formulae with bad smells~\cite{hermans2015detecting}. 

One of the well-recognized spreadsheet smells is \nestifexp{}~\cite{Hermans2015,hermans2015detecting}. \emph{IF} functions\footnote{Functions are predefined built-in formulae already available in spreadsheet systems. } (i.e., the syntax is $IF(condition, true\_branch, \\false\_branch)$) are widely used spreadsheet functions. \emph{Nested-IF} expressions happen when end users write an \emph{IF} function inside another \emph{IF} or \nestif{} function. According to previous research~\cite{6976077,7476773,Tufano:2015,Hermans2015,hermans2015detecting}, \nestifformu{} in speadsheets are complex, unreadable, error-prone, as well as hard to debug and maintain. There are also a lot of online discussions about the harm of \nestifformu{}. Some people have expressed their desire to reduce \nestif{}s ``wherever possible''~\cite{19_tips_for_nested_if_formulas,reddit2015,reddit_neverusenf}. 

What is worse, the bad practice of using \nestifexp{} among end users is quite common: our study of over 68,000 real-world spreadsheets\footnote{In this paper, we refer spreadsheet as a file consisting of one or multiple worksheets~\cite{hermans2015enron}.} reveals that for the worksheets containing \emph{IF}, 30.04\% of them also contain \nestif{}. If we denote the maximum nesting level inside a \nestif{} as \emph{if-depth} \footnote{E.g, $IF(IF(L1>=F\$5,L1),IF(L1<=F\$6,L1,``"),``")$ has an \emph{if-depth} of 2.}, in our corpus each spreadsheet includes on average 9 formulae with \ifdepth{} over 10, while the observed maximum \emph{if-depth} is 48 with multiple instances.

Formula refactoring is a practical solution to tackle this problem, which was first proposed by Badame and Dig~\cite{badame2012refactoring}: to perform semantic-preserving formula transformations (without changing the behavior) with the purpose of removing formulae smells. Nevertheless, such refactoring requires essential knowledge and skills of programming which is challenging for end users. To help end users, several previous works~\cite{badame2012refactoring,Hermans2015,hermans2014bumblebee} have proposed a few simple refactoring patterns trying to decrease the \ifdepth{}, but they either have very low coverage (i.e., the ratio of formulae that can be ameliorated) or are non-automatic.

In this paper, we firstly propose an AST (Abstract Syntax Tree) based approach to systematically tackling this problem via automated refactoring. The general idea is two-fold. First, there often exists logic redundancy across different condition paths within a \nestif{}. Reduction of the redundant logic can remove useless parts and simplify the \nestif{} formula. Second, some higher-level semantics are often fragmented into hierarchical combinations of \emph{IF} conditions in a \nestif{}. Reassembling the fragmented syntax from corresponding \emph{IF}-subtrees into built-in functions can shorten the \nestif{} formula. To analyze and refactor both redundant logic and fragmented syntax, our approach leverages and works on the AST (Abstract Syntax Tree) structure as intermediate representation of \nestif{} formulae.

The evaluation is conducted on over 68,000 real-world spreadsheets with over 27 million \nestifformu{}. The experimental results lead to the following three key takeaways. First, our approach is generally applicable - over 99\% of the \nestifformu{} can be refactored and the refactor has been verified as correct. Second, our approach is effective - over 50\% of the refactoring have achieved more than half of their \ifdepth{} reduced; while the \nestiffunc{} in most formulae are completely reduced or transformed with \ifdepth{} 1. Third, end users recognize our approach and its results. A survey on 49 participants indicates that most of them prefer the refactored formulae and believe the automated refactoring is necessary and helpful; while only a few of them are equipped with the knowledge of manual refactoring.

The main contributions of this paper are shown as follows.

\noindent \textbf{1) An automated and highly-effective approach to identify and refactor \nestifformu{}.} The goal is to help end users reduce the complexity and cognitive cost of \nestifformu{} in spreadsheets. 

\noindent \textbf{2) A comprehensive evaluation of the proposed automated approach.} We evaluated the correctness, applicability, effectiveness and usefulness of the approach.

\noindent \textbf{3) A statistical study on the current usage of \nestifformu{} in real-world spreadsheets.} We present detailed statistics of \nestifformu{} against over 68,000 real-world  spreadsheets collected in a leading IT company for research purpose.



\section{Preliminary Study}
\label{sec:premilinarystudy}
Our preliminary study aims to present the usage status of \nestifformu{} and to further motivate our approach.

The investigation is based on a spreadsheet corpus including over 68,000 real-world Excel files (a.k.a. spreadsheets/workbooks), which contain a total of over 149,170 worksheets. The source of this corpus is a spreadsheet repository collected in a leading IT company for research purposes. And the files in our corpus are extracted by excluding those with technical complications as obstacles for interaction-free processing (e.g., password protected, external reference embedded requiring trust confirmation). Compared to the corpora that have been widely used in previous research work--the EUSES Spreadsheet Corpus~\cite{fisher2005euses}, Enron Spreadsheet Corpus~\cite{hermans2015enron}, and Hawaii Kooker Corpus~\cite{aurigemma2010detection}--our corpus has the following two advantages:


\noindent \textbf{1) Larger scale.} The number of spreadsheets (68,075) is much larger than the Enron corpus (15,770), EUSES corpus (4,037), and Hawaii Kooker Corpus (74). Table~\ref{tab:corpus} lists a detailed comparison between our corpus and the previous largest corpus Enron\footnote{We do not compare with the other two corpora because their file type is too old for Python excel library \emph{openpyxl}~\cite{gazoni2016openpyxl} to parse.}.

\noindent \textbf{2) Higher diversity regarding domains.} The corpus contains diverse spreadsheets for various purposes across multiple domains, while the other corpora either contain large numbers of toy spreadsheets or come from a single company of a specific domain.


\begin{table}[h]\small
	\vspace{-2mm}
	\caption{\label{tab:corpus} Comparison between Enron and our corpus}
	\begin{tabular}{l|r|r}
		\toprule
		&\multicolumn{2}{c}{\textbf{Corpus}}\\ \cline{2-3}
		&\textbf{Enron}&\textbf{Ours}\\ \midrule 
		Total number of spreadsheets&15,770&68,075 \\ 
		Total number of worksheets&79,983&149,170 \\ 
		Average size of spreadsheets&113.4~KB&1,211.3~KB \\ 
		Number of spreadsheets with formulae&9,120&37,109\\ 
		Number of spreadsheets with \emph{IF} functions&2,020&14,425\\ 
		Number of \emph{IF} functions&3,420,790&138,085,568\\ \bottomrule
		
	\end{tabular}
	\vspace{-2mm}	
\end{table}

Based on this very large and diverse corpus, we investigate the usage status of \nestifformu{}. Inside one spreadsheet many formulae may be created by dragging one formula down or to the right to repeat its calculation. As in previous work~\cite{hermans2015enron,7739679,jansen2015enron}, we remove these formulae by clustering the formulae based on their R1C1 notation\footnote{The R1C1 notation will stay the same even if the formula is dragged down or right.}. We then pick one formula from each cluster to form the new formula set. We call this new set the ``Unique Set'' and the original set the ``Total Set''.

The results are shown in Table~\ref{tab:formuanum}. The first two rows show the results of the formula number. The remaining rows show the number of formulae in different depth ranges. Column ``Total''/``Unique'' shows the results of the Total/Unique Set, From the table, among the 27,689,699 total formulae and 19,260,407 unique formulae, over 12\% formulae have an \ifdepth{} of over 5, over 75,000 total formulae and 35,000 unique formulae even have an \ifdepth{} of over 15, indicating surprisingly heavy usage of \nestifformu{}. 

\begin{table}[h]\small
		\vspace{-2mm}
	\caption{\label{tab:formuanum}Usage status of \nestifformu{}}
	\begin{tabular}{p{4.7cm}|r|r} 
		\toprule
		&\textbf{Total}&\textbf{Unique} \\ \midrule 
		Formula number&27,689,699&19,260,407 \\ 
		Formula number per-spreadsheet&407&283 \\ 
		\ifdepth{} in range (1,5]&24,206,022&16,680,744	\\ 
		\ifdepth{} in range (5,10]&2,815,521&2,408,549 \\ 	
		\ifdepth{} in range (10,15]&548,129&118,355 \\ 
		\ifdepth{} in range (15,65]&75,455& 35,201\\ \bottomrule	
	\end{tabular}
	\vspace{-2mm}
\end{table}

This heavy usage of common \nestifformu{} would cause much harm as we mentioned in the introduction. However, end-users usually lack the awareness of such harm. They also tend to lack enough spreadsheet function knowledge to manually refactor or avoid using \nestifformu{}. Consequently, it is essential that automatic approaches should be constructed to help tackle these problems.

After manually checking a sample of these \nestifformu{}\footnote{The authors manually checked the \nestifformu{} in 100 randomly-sampled spreadsheets.}, we realize that many formulae contain unnecessary conditions which would cause dead branches\footnote{A dead branch will never be executed.}. Additionally, a large part of \nestiffunc{} actually combine together to fulfill a certain functionality, which can also be fulfilled by other high-level functions already defined in spreadsheets like Excel. For example, the following formula is a real case from our corpus: {\tt $IF(IF(Q1=X1,Q1,IF(Q1=``",X1,IF(Q1<>X1,Q1)))=``",``",IF(Q1=X1,\\Q1,IF(Q1=``",X1,IF(Q1<>X1))))$}. If we use $M$ to represent sub-string $IF(Q1=X1,Q1,IF(Q1=``",X1,IF(Q1<>X1,Q1)))$, then the formula can be written as $IF(M=``",``",M)$. To better illustrate the formula structure, in Figure~\ref{fig:astexample} we present the general AST in the left as well as the AST of $M$ in the right. From Figure~\ref{fig:astexample}, it is easy to tell that condition $Q1<>X1$ is redundant: this condition is in the false branch of condition $Q1=X1$, and thus is certain to be true. Additionally, $IF(M=``",``",M)$ actually equals to $M$ no matter what value $M$ is. If we remove these unnecessary \emph{IF} expressions, the formula can be refactored into $IF(Q1=X1,Q1,IF(Q1=``",X1,Q1)$. Furthermore, we can use the IFS function to transform $IF(Q1=X1,Q1,IF(Q1=``",X1,Q1)$ into $IFS(Q1=X1,Q1,Q1=``'',X1,TRUE,Q1)$, which is much cleaner than the original one. Inspired by this observation, we propose our AST-based approach to automatically accomplishing such kind of formula refactorings. 

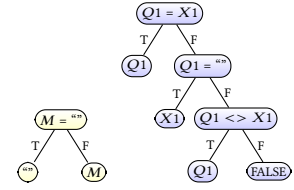
\begin{figure}[t]\tiny
	\centering
	\begin{tabular}{cc}	
\begin{tikzpicture}
[sibling distance=5em,
level distance=0.7cm,
treenode/.style = {shape=rectangle, rounded corners,
	draw, align=center,
	top color=white, bottom color=yellow!20},
payoff/.style    = {align=center, inner sep=0.1em, text width=1.5em},
left side node/.style={above left, inner sep=0.1em},
right side node/.style={above right, inner sep=0.1em}]
\node[treenode]{$M=``"$}
child {node [treenode] {$``"$}edge from parent node[left] {T}}
child {node[treenode] {$M$} edge from parent node[right] {F} 
	child[missing] {}
	child[missing] {}
};\end{tikzpicture} &\hspace{-2mm}
	\begin{tikzpicture}
[sibling distance=5em,
level distance=0.7cm,
treenode/.style = {shape=rectangle, rounded corners,
	draw, align=center,
	top color=white, bottom color=blue!20},
payoff/.style    = {align=center, inner sep=0.1em, text width=1.5em},
left side node/.style={above left, inner sep=0.1em},
right side node/.style={above right, inner sep=0.1em}]
	\node [treenode]{$Q1=X1$}
	child {node [treenode] {$Q1$}edge from parent node[left] {T}}
	child {node [treenode] {$Q1=``"$}	
		child {node  [treenode]{$X1$}edge from parent node[left] {T}}
			child {node [treenode] {$Q1<>X1$}
				child {node [treenode] {$Q1$}edge from parent node[left] {T}}
						child {node [treenode] {FALSE}edge from parent node[right] {F}}	edge from parent node[right] {F}
					}
					edge from parent node[right] {F}
			};	
		\end{tikzpicture}
\end{tabular}
\caption{AST example} \label{fig:astexample}
\vspace{-4mm}
\end{figure}

\section{Approach}
\label{sec:approach}
\subsection{Overview}

By analyzing the AST structure of each formula, our approach identifies optimizable \nestifexp{} and performs refactoring by replacing basic-level and counter-intuitive syntax with non-redundant and high-level syntax. The major rationale behind using AST is the desirable structural mapping between AST and \nestif{} as follows. An \emph{IF} function typically contains three parts: 1) condition, 2) true-branch expression, and 3) false-branch expression. Therefore, the ASTs of \nestifexp{} are binary trees, with the true- and false-branch expressions being the two child-nodes of the condition node. Consequently, with AST, it is easy to detect and locate \nestif{} in a formula as well as convenient to conduct further analysis based on the tree structure.



In this section, we first introduce a high-level overview of our 3-step algorithm framework, followed by detailed introductions to the two key algorithms for redundancy removal and syntax reassembling, respectively.

 \textbf{Step1: AST generation.} We parse each formula and generate its AST to support the subsequent analysis. AST is a tree representation of the abstract syntactic structure of source code written in a programming language. In spreadsheet related research, AST is usually adopted to indicate formula complexity~\cite{reschenhofer2017conceptual}. The larger depth (height) of the AST, the higher complexity of the formula.

Based on the AST of the formula, we then traverse the AST and calculate the \ifdepth{}. Specifically, along each path of AST, we record the number of \emph{IF} functions, and regard the largest one across all paths as the \ifdepth{} of the formula. For example, the \ifdepth{} of formula $IF(condition1,IF(condition2,value1,value2), IF(condition3,IF(\\condition4,value3,value4),value5)$ is 3.

A \nestif{} is identified in a formula when its \ifdepth{} is greater than 1, and will be passed to the subsequent steps for refactoring analysis. Otherwise, if the \ifdepth{} equals 0 (i.e., no \emph{IF} in this formula) or 1 (i.e., no \nestif{} in this formula), our algorithm will bypass the formula directly.

 \textbf{Step2: Redundancy removal.} An \emph{IF} expression can essentially be mapped to an if-else branching statement in professional programming. Once the condition on some node remains deterministic due to its preceding evaluation at some ancestor node on AST, it will become a redundant condition and one of its child branches must be dead code. Such redundant conditions are spreadsheet smells that require removal, since they introduce unnecessary complications to the spreadsheet data thus may confuse end users. We conduct such redundancy removal first, because its existence may also obscure the AST structure from well understood patterns and thus put negative impact on our pattern matching for syntax reassembling. More details of this step can be found in Section~\ref{sec:redunremove}.

 \textbf{Step3: Syntax reassembling.} We have observed another typical smell in real-world spreadsheets, where single and higher-level semantics are often fragmented by end user into lower-level syntax pieces with \nestif{}s. In fact, for such semantics there are concise and easily understood forms in spreadsheet systems with built-in functions. The goal of this step is to conduct reverse inference against such a smell, i.e., to recognize and reassemble such semantic-fragmented AST regions into their more concise forms via pattern matching and replacement. In this paper, we have manually identified 9 patterns for 9 major types of semantics respectively. These patterns are summarized based on our case analysis. First, we sampled around 100 spreadsheets from the large-scale corpus. Second, we manually studied the samples and came up with the patterns by summarization and abstraction, combining with our own knowledge. Each pattern corresponds to a spreadsheet built-in functions\footnote{Most mainstream spreadsheet tools such as Excel and Google Sheets support these functions.}. We present the name, explanation, and examples of each alternative function in Table~\ref{tab:ninefunc}. As of the composing of this paper, there might be other function candidates that remain out of our knowledge. Nonetheless, our proposed algorithm framework should be extensible for easy incorporation of new patterns, and we do plan to continue related study in the future accordingly. More details of this step can be found in Section~\ref{sec:alterreplace}.

\subsection{Redundancy Removal}
\label{sec:redunremove}

In this section, we introduce how we identify and remove redundant conditions in a formula (Step 2). The procedure is presented in Algorithm~\ref{al:removeredun}, with the help of an example flow in Figure~\ref{fig:redunprocess}.

\begin{algorithm} [!h]\small
	\KwIn{$F_m$: the current nested-IF formula }
	\KwIn{$AST$: the AST of $F_m$ }
	\KwOut{$F_m$: the new formula}
	containRedun $\leftarrow$ TRUE\\
	\While{containRedun}{
		containRedun $\leftarrow$ FALSE\\
		$ifList$ $\leftarrow$ getParentIfList($AST$, $F_m$)\\	
		\For{each $ifexp$ in $ifList$}
		{
			$ifTree$ = generateAST($ifexp$)\\
			dicConBranch $\leftarrow$ getDicConBranch($ifTree$)\\
			\For{each condition in dicConBranch}
			{
				branchList $\leftarrow$ dicConBranch.get(condition)\\
				dBranchList $\leftarrow$ branchfList.directpart\\
				nBranchList$\leftarrow$ branchfList.negativepart\\
				\If{dBranchfList.len $>$ 2}
				{				
					containRedun $\leftarrow$ TRUE \tcp{Redundancy exists.}
					redunIFList $\leftarrow$ generateRedunList(condition, dBranchList)\\
					\For{each redunIF in redunIFList }
					{
						\eIf{redunIF in condition.truebranch}
						{$F_m$ $\leftarrow$ $F_m$.replace(redunIF, redunIF.truebranch)}
						{$F_m$ $\leftarrow$ $F_m$.replace(redunIF, redunIF.falsebranch)}}}
				\If{nBranchList.len $>$ 0}
				{
					containRedun $\leftarrow$ TRUE \tcp{Redundancy exists.}
					redunIFList $\leftarrow$ generateRedunList(condition, nBranchList)\\
					\For{each redunIF in redunIFList }
					{
						\eIf{redunIF in condition.falsebranch}
						{$F_m$ $\leftarrow$ $F_m$.replace(redunIF, redunIF.truebranch)}
						{$F_m$ $\leftarrow$ $F_m$.replace(redunIF, redunIF.falsebranch)}}
				}
				$AST$ $\leftarrow$ generateAST($F_m$)
			}
		}
	}
	return $F_m$
	
	\caption{Function: removeRedun}
	\label{al:removeredun}
	
\end{algorithm}


\textbf{1) \emph{Nested-IF} expression extraction}. First, we extract outmost \nestifexp{} from each formula (see Line 4 in Algorithm~\ref{al:removeredun}, function \emph{getParentIfList}). By outmost we mean the highest hierarchy in a nested branching logic or on an AST. For example, as shown in Figure~\ref{fig:redunprocess}, for formula $SUM(IF(C1, V1, IF(!C1,V2,IF(C2,\\V3,V4))),V5)$, there is only one target \emph{nested-IF} expression: the outmost IF expression: $IF(C1, V1, IF(!C1,V2,IF(C2,V3,V4)))$; for formula $SUM(IF(C1,V1,V2),IF(C2,V3,IF(!C2,V4,V5)),IF(C3,V5,\\IF(IF(C4,V6,V7))))$, there are two target \nestifexp{}: $IF(C2,V3,IF(!C2,V4,V5))$ and $IF(C3,V5,IF(IF(C4,V6,V7)))$. Please note that the \nestif{} targets to be analyzed may also exist in predicates of a condition node, and we also extract such \emph{IF} expressions. For example, $IF(C1,V1,IF(!C1,V2,V3))$ is also extracted from the condition part of formula $IF(IF(C1,V1,IF(!C1,V2,V3))=V1, V3, IF(!C1,V4,V5))$. By doing so, we do not miss the chance of optimizing the \nestif{} at condition parts. Moreover, in case the \nestif{} would be reduced as simple predicates, it would potentially increase the chance of eventually optimizing the outmost \nestif{}.

\textbf{2) Branch collection}. Based on the AST of each extracted \nestif{}, we create a dictionary \emph{dicConBranch} as the key structure to help detect and remove redundant logic (function $getDicConBranch$ in Line 7). As shown in Figure~\ref{fig:redunprocess}, for each entry in the dictionary, its key is the condition of an AST node such as $C1$ or $C2$; the $dBranchList$ value (Line 10) stores a tuple of two AST sub-trees corresponding to true and false branches respectively. In addition, each entry also has a $nBranchList$ value (Line 11) for the negation of the key condition such as $!C1$, and stores the tuple of true and false branches accordingly. The dictionary is constructed by visiting each condition node on the AST. When the same condition (or negation) is hit for multiple times, the AST sub-tree tuples at each hitting site are appended to the $dBranchList$ (or $nBranchList$).

\textbf{3) Redundancy identification and removal}. Intuitively, if any entry stores more than 1 tuple in $dBranchList$ and $nBranchList$ collectively (Line 12, 24), it indicates existence of redundant branches on the AST about the condition at key. We iterate such inspection against \emph{dicConBranch} to detect and remove redundancies. Each detected redundancy site corresponds to one redundant \emph{IF} expression that can be replaced with either the true branch (the condition is deterministic as true) or the false branch (the condition is deterministic as false). Thus, under each situation, we generate the redundant \emph{IF} expression according to the condition and its branch list (Lines 15 and 26) and make replacement. For example, as the example in Figure~\ref{fig:redunprocess} shows, the \emph{nBranchList} of key $C1$ is not null, indicating that \emph{IF} expression $IF(!C1,V2,IF(C2,V3,V4))$ is redundant. Since this expression lies in the false branch of condition $C1$, condition $!C$ is deterministic as true. Therefore, we remove condition $!C1$ and its false branch, and only keep the true branch $V2$. As a result, the original formula becomes $SUM(IF(C1, V1, V2),V5)$. Such iteration repeats until no redundancy is detected.

\begin{figure*}[h]\tiny
	\begin{tabular}{ccccccccccc}	
		\includegraphics[scale=0.3]{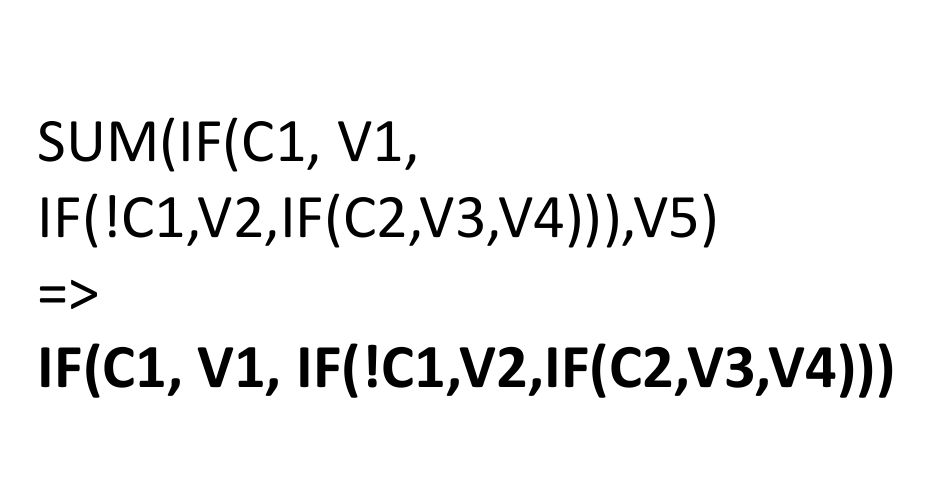}
		&\hspace{-6mm}
		\includegraphics[scale=0.3]{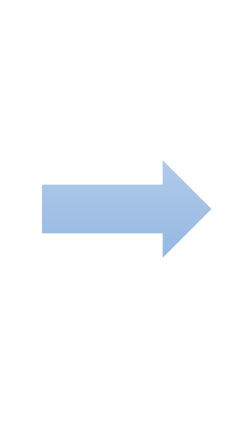}&\hspace{-4mm}
		\includegraphics[scale=0.2]{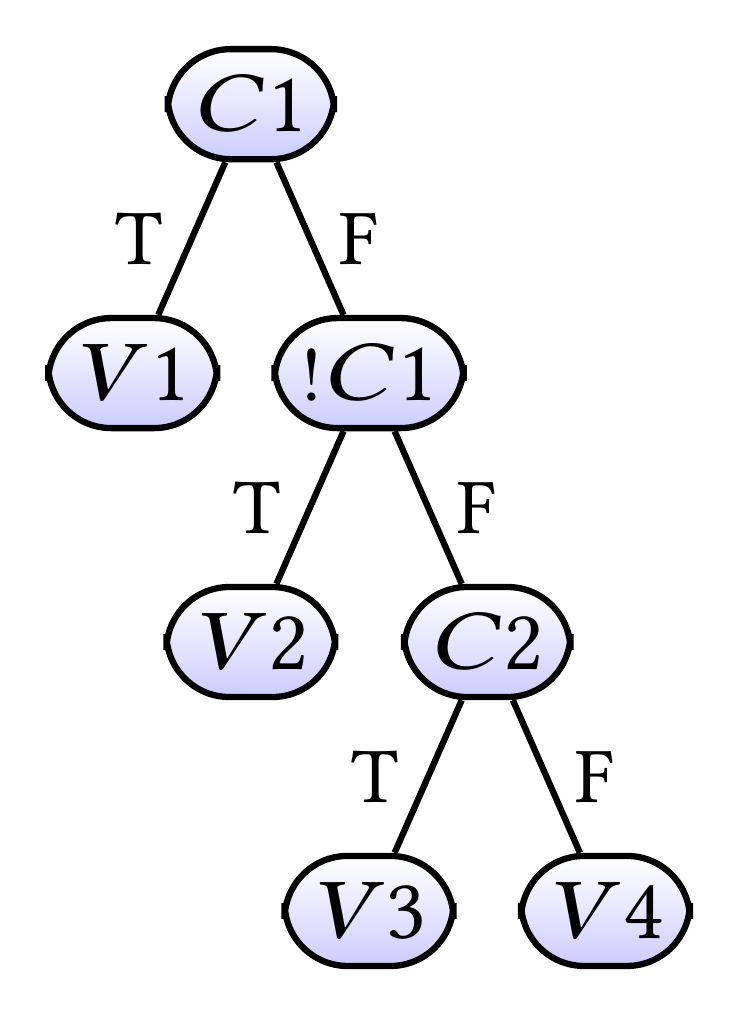}&\hspace{-4mm}
		\includegraphics[scale=0.3]{figures/arrow}&\hspace{-4mm}
		\includegraphics[scale=0.25]{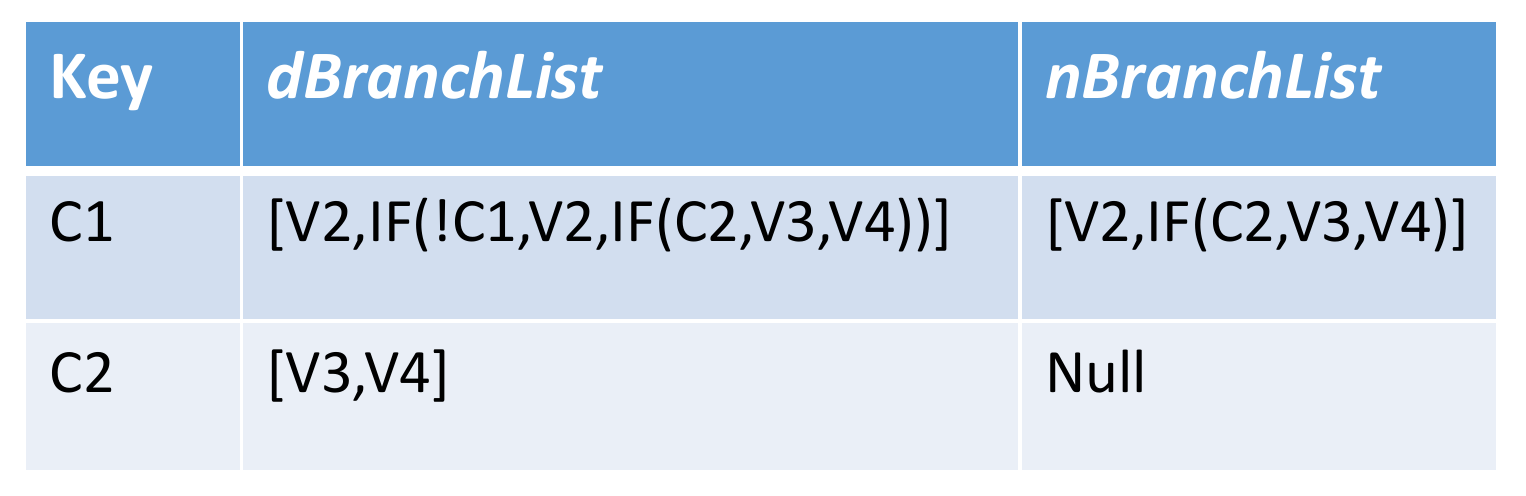}&\hspace{-4mm}
		\includegraphics[scale=0.3]{figures/arrow}&\hspace{-4mm}
		\includegraphics[scale=0.2]{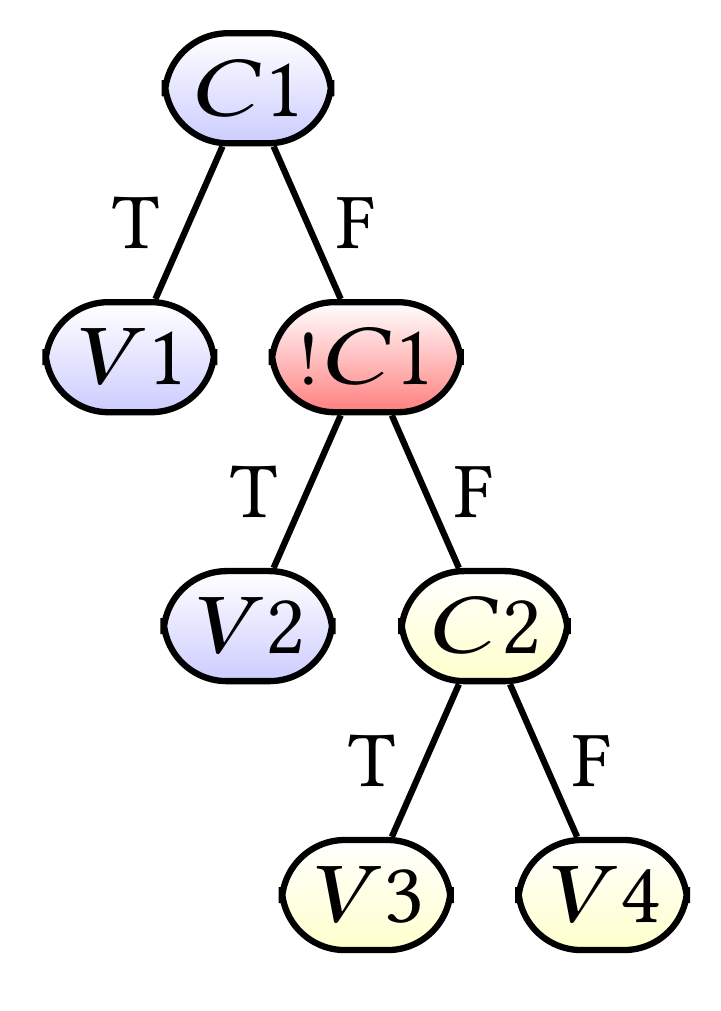}&\hspace{-4mm}
		\includegraphics[scale=0.3]{figures/arrow}&\hspace{-4mm}
		\includegraphics[scale=0.2]{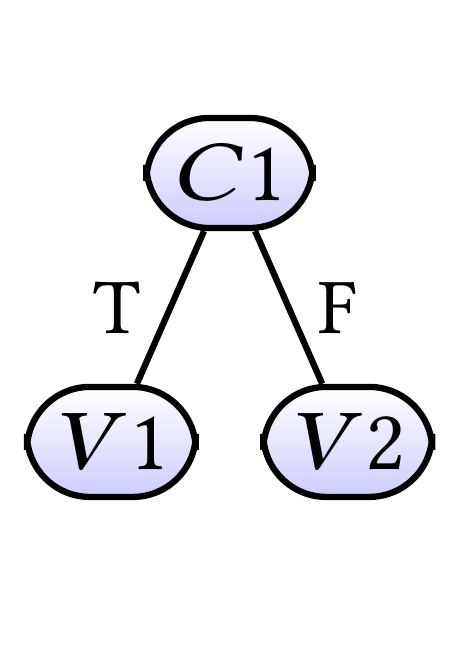}&
		\includegraphics[scale=0.3]{figures/arrow}&\hspace{-4mm}
		\includegraphics[scale=0.2]{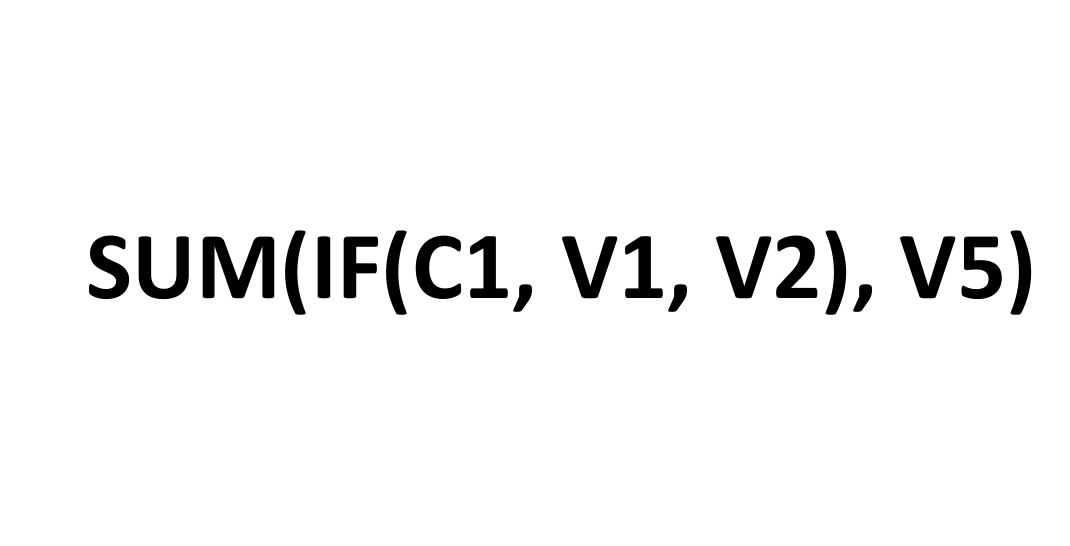}\\
	\end{tabular}
	\caption{Redundancy removal process.}\label{fig:redunprocess}
\end{figure*}
\subsection{Syntax Reassembling}
\label{sec:alterreplace}

After removing redundancies, if the resultant formula still contains \nestifexp{}, in this third step we further analyze the AST to detect and reassemble fragmented semantics into built-in functions as listed in Table~\ref{tab:ninefunc}.

\subsubsection{General Procedure}
In general, this step is in a paradigm of iterative pattern-matching and replacement. For each remaining \nestif{} after step 2, we further construct a $threePartList$ as the key structure to facilitate pattern matching. Each $threePartList$ consists of three lists for condition, true branch, and false branch, respectively. For example, for expression $IF(C1,IF(C2,IF(C3,V1,V2),\\V2),V2)$, the condition part is $[C1,C2,C3]$, the true branch part is $[IF(C2,IF(C3,V1,V2),V2), IF(C3,V1,V2),V1]$, and the false part is $[V2,V2,V2]$.

Subsequently, based on \emph{threePartList}, we infer the semantic of the \emph{IF} expression and check if it matches some spreadsheet functions. If yes, we transform the formula using the matched function, and replace the \nestif{} expression with the transformed one. Following the order shown in Table~\ref{tab:ninefunc}, we probe each pattern in sequence. Once a pattern is matched, the probe jumps to the next iteration from the first pattern again. This iteration terminates with zero pattern match. Note that the patterns $CHOOSE$/$MATCH$/$LOOKUP$ have higher priority than the pattern $IFS$ during the matching, because they are more comprehensible and enable more concise expressions. In the future, we may consider to provide all alternative refactoring recommendations for end users to choose from.
\begin{figure*}[h!]\tiny
	\centering
	\begin{tabular}{cccccccc}		
	\begin{tikzpicture}
		[sibling distance=3em,
		level distance=0.6cm,
		treenode/.style = {shape=rectangle, rounded corners,draw, align=center,top color=white, bottom color=blue!20},
		leaveonenode/.style = {shape=rectangle, rounded corners,draw, align=center,top color=white, bottom color=yellow!20},
		leavetwonode/.style = {shape=rectangle, rounded corners,draw, align=center,top color=white, bottom color=green!20},
		payoff/.style    = {align=center, inner sep=0.1em, text width=1.5em},
		left side node/.style={above left, inner sep=0.1em},
		right side node/.style={above right, inner sep=0.1em}]
	
		\node[treenode]{$C1$}
		child {node [treenode] {$C2$} 
			child {node [treenode] {$C3$}
				child {node [treenode] {$C4$}
					child {node[leaveonenode] {$V1$}edge from parent node[left] {T}}
					child {node[leavetwonode] {$V2$}edge from parent node[right] {F}}
					edge from parent node[left] {T}}
				child {node[leavetwonode] {$V2$}edge from parent node[right] {F} } 
				edge from parent node[left] {T}	}
			child {node[leavetwonode] {$V2$}edge from parent node[right] {F} } 
			edge from parent node[left] {T}	}
		child {node[leavetwonode] {$V2$} edge from parent node[right] {F}
			child[missing] {}
			child[missing] {}
		};
	\node[above,font=\bfseries] at (current bounding box.north) {AND};
	\end{tikzpicture} &
		\begin{tikzpicture}
		[sibling distance=3em,
		level distance=0.6cm,
		treenode/.style = {shape=rectangle, rounded corners,draw, align=center,top color=white, bottom color=blue!20},
		leaveonenode/.style = {shape=rectangle, rounded corners,draw, align=center,top color=white, bottom color=yellow!20},
		leavetwonode/.style = {shape=rectangle, rounded corners,draw, align=center,top color=white, bottom color=green!20},
		payoff/.style    = {align=center, inner sep=0.1em, text width=1.5em},
		left side node/.style={above left, inner sep=0.1em},
		right side node/.style={above right, inner sep=0.1em}]
		\node [treenode]{$C1$}
		child {node [leaveonenode] {$V1$}edge from parent node[left] {T}}
		child {node [treenode] {$C2$}	
			child {node  [leaveonenode]{$V1$}edge from parent node[left] {T}}
			child {node [treenode] {$C3$}
				child {node [leaveonenode] {$V1$}edge from parent node[left] {T}}
				child {node [treenode] {$C4$}
				child {node [leaveonenode] {$V1$}edge from parent node[left] {T}}
					child {node [leavetwonode] {$V2$}edge from parent node[right] {F}}edge from parent node[right] {F}}	edge from parent node[right] {F}
			}
			edge from parent node[right] {F}
		};	
	\node[above,font=\bfseries] at (current bounding box.north) {OR};
		\end{tikzpicture}&\hspace{-5mm}
		
		\begin{tikzpicture}
		[sibling distance=5em,
		level distance=0.6cm,
		treenode/.style = {shape=rectangle, rounded corners,draw, align=center,top color=white, bottom color=blue!20},
		leaveonenode/.style = {shape=rectangle, rounded corners,draw, align=center,top color=white, bottom color=yellow!20},
		leavetwonode/.style = {shape=rectangle, rounded corners,draw, align=center,top color=white, bottom color=green!20},
		payoff/.style    = {align=center, inner sep=0.1em, text width=1.5em},
		left side node/.style={above left, inner sep=0.1em},
		right side node/.style={above right, inner sep=0.1em}]
		\node [treenode]{$A1=n1$}
		child {node [leaveonenode] {$str1$}edge from parent node[left] {T}}
		child {node [treenode] {$A1=n2$}	
			child {node  [leaveonenode]{$str2$}edge from parent node[left] {T}}
			child {node [treenode] {$A1=n3$}
				child {node [leaveonenode] {$str3$}edge from parent node[left] {T}}
				child {node [treenode] {$A1=n4$}
					child {node [leaveonenode] {$str4$}edge from parent node[left] {T}}
					child {node [leaveonenode] {$FALSE$}edge from parent node[right] {F}}edge from parent node[right] {F}}	edge from parent node[right] {F}
			}
			edge from parent node[right] {F}
		};	
		\node[above,font=\bfseries] at (current bounding box.north) {CHOOSE};
		\end{tikzpicture} &\hspace{-10mm}
	\begin{tikzpicture}
	[sibling distance=5em,
	level distance=0.6cm,
	treenode/.style = {shape=rectangle, rounded corners,draw, align=center,top color=white, bottom color=blue!20},
	leaveonenode/.style = {shape=rectangle, rounded corners,draw, align=center,top color=white, bottom color=yellow!20},
	leavetwonode/.style = {shape=rectangle, rounded corners,draw, align=center,top color=white, bottom color=green!20},
	payoff/.style    = {align=center, inner sep=0.1em, text width=1.5em},
	left side node/.style={above left, inner sep=0.1em},
	right side node/.style={above right, inner sep=0.1em}]
	\node [treenode]{$A1=str1$}
	child {node [leaveonenode] {$n1$}edge from parent node[left] {T}}
	child {node [treenode] {$A1=str2$}	
		child {node  [leaveonenode]{$n2$}edge from parent node[left] {T}}
		child {node [treenode] {$A1=str3$}
			child {node [leaveonenode] {$n3$}edge from parent node[left] {T}}
			child {node [treenode] {$A1=str4$}
				child {node [leaveonenode] {$n4$}edge from parent node[left] {T}}
				child {node [leaveonenode] {$FALSE$}edge from parent node[right] {F}}edge from parent node[right] {F}}	edge from parent node[right] {F}
		}
		edge from parent node[right] {F}
	};	
	\node[above,font=\bfseries] at (current bounding box.north) {MATCH};
	\end{tikzpicture} &\hspace{-10mm}
		\begin{tikzpicture}
		[sibling distance=5em,
		level distance=0.6cm,
		treenode/.style = {shape=rectangle, rounded corners,draw, align=center,top color=white, bottom color=blue!20},
		leaveonenode/.style = {shape=rectangle, rounded corners,draw, align=center,top color=white, bottom color=yellow!20},
		leavetwonode/.style = {shape=rectangle, rounded corners,draw, align=center,top color=white, bottom color=green!20},
		payoff/.style    = {align=center, inner sep=0.1em, text width=1.5em},
		left side node/.style={above left, inner sep=0.1em},
		right side node/.style={above right, inner sep=0.1em}]
		\node [treenode]{$A1=r1$}
		child {node [leaveonenode] {$r2$}edge from parent node[left] {T}}
		child {node [treenode] {$A1=r3$}	
			child {node  [leaveonenode]{$r4$}edge from parent node[left] {T}}
			child {node [treenode] {$A1=r5$}
				child {node [leaveonenode] {$r6$}edge from parent node[left] {T}}
				child {node [treenode] {$A1=r7$}
					child {node [leaveonenode] {$r8$}edge from parent node[left] {T}}
					child {node [leaveonenode] {$FALSE$}edge from parent node[right] {F}}edge from parent node[right] {F}}	edge from parent node[right] {F}
			}
			edge from parent node[right] {F}
		};	
		\node[above,font=\bfseries] at (current bounding box.north) {LOOKUP};
		\end{tikzpicture}&\hspace{-1mm}
		\begin{tikzpicture}
		[sibling distance=5em,
		level distance=0.6cm,
		treenode/.style = {shape=rectangle, rounded corners,draw, align=center,top color=white, bottom color=blue!20},
		leaveonenode/.style = {shape=rectangle, rounded corners,draw, align=center,top color=white, bottom color=yellow!20},
		leavetwonode/.style = {shape=rectangle, rounded corners,draw, align=center,top color=white, bottom color=green!20},
		payoff/.style    = {align=center, inner sep=0.1em, text width=1.5em},
		left side node/.style={above left, inner sep=0.1em},
		right side node/.style={above right, inner sep=0.1em}]
		\node[treenode]{$A>B$}
		child {node [leaveonenode] {$A$}edge from parent node[left] {T}}
		child {node[leaveonenode] {$B$} edge from parent node[right] {F} 
			child[missing] {}
			child[missing] {}
		};
	\node[above,font=\bfseries] at (current bounding box.north) {MAX};
	\end{tikzpicture}&\hspace{-1mm}
	\begin{tikzpicture}
	[sibling distance=5em,
	level distance=0.6cm,
	treenode/.style = {shape=rectangle, rounded corners,draw, align=center,top color=white, bottom color=blue!20},
	leaveonenode/.style = {shape=rectangle, rounded corners,draw, align=center,top color=white, bottom color=yellow!20},
	leavetwonode/.style = {shape=rectangle, rounded corners,draw, align=center,top color=white, bottom color=green!20},
	payoff/.style    = {align=center, inner sep=0.1em, text width=1.5em},
	left side node/.style={above left, inner sep=0.1em},
	right side node/.style={above right, inner sep=0.1em}]
	\node[treenode]{$A<B$}
	child {node [leaveonenode] {$A$}edge from parent node[left] {T}}
	child {node[leaveonenode] {$B$} edge from parent node[right] {F} 
		child[missing] {}
		child[missing] {}
	};
\node[above,font=\bfseries] at (current bounding box.north) {MIN};
\end{tikzpicture}&\hspace{-5mm}
		\begin{tikzpicture}
		[sibling distance=3em,
		level distance=0.6cm,
		treenode/.style = {shape=rectangle, rounded corners,draw, align=center,top color=white, bottom color=blue!20},
		leaveonenode/.style = {shape=rectangle, rounded corners,draw, align=center,top color=white, bottom color=yellow!20},
		leavetwonode/.style = {shape=rectangle, rounded corners,draw, align=center,top color=white, bottom color=green!20},
		payoff/.style    = {align=center, inner sep=0.1em, text width=1.5em},
		left side node/.style={above left, inner sep=0.1em},
		right side node/.style={above right, inner sep=0.1em}]
		\node [treenode]{$C1$}
		child {node [leaveonenode] {$V1$}edge from parent node[left] {T}}
		child {node [treenode] {$C2$}	
			child {node  [leaveonenode]{$V2$}edge from parent node[left] {T}}
			child {node [treenode] {$C3$}
				child {node [leaveonenode] {$V3$}edge from parent node[left] {T}}
				child {node [treenode] {$C4$}
					child {node [leaveonenode] {$V4$}edge from parent node[left] {T}}
					child {node [leaveonenode] {$V5$}edge from parent node[right] {F}}edge from parent node[right] {F}}	edge from parent node[right] {F}
			}
			edge from parent node[right] {F}
		};	
		\node[above,font=\bfseries] at (current bounding box.north) {IFS};
		\end{tikzpicture}

	\end{tabular}
	\caption{Typical AST of function AND,OR,CHOOSE,MATCH,LOOKUP, MAX,MIN, and IFS. $stri$ represents a string; $ni$ represents a number; $ri$ represents a reference ($0<i<5$). } \label{fig:typicalast}
\end{figure*}
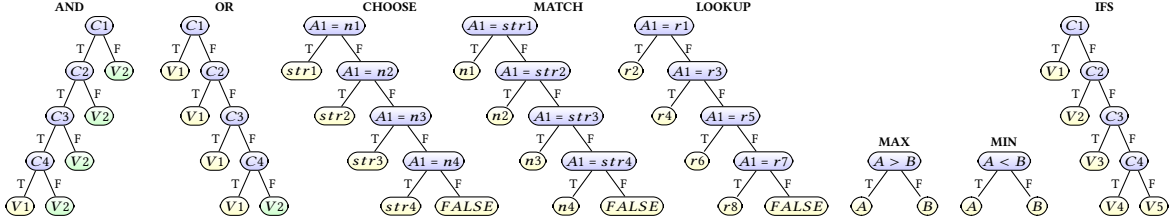

\subsubsection{Alternative Functions}

 
 \begin{table*}[!h]\small
 	\center
 	\caption{\label{tab:ninefunc}The advanced functions}
 	\vspace{-1mm}
 	\begin{tabular}{  p{1.1cm} | p{6.8cm} |p{8.8cm}}
 		\toprule
 		\textbf{Name} & 	\textbf{Explanation} & 	\textbf{Transformation Examples }\\ \midrule 	
 		AND&Returns TRUE if all of the arguments evaluate to TRUE.&$IF(C1,IF(C2,IF(C3,V1,V2),V2),V2)$ $\rightarrow$ $IF(AND(C1,C2,C3),V1,V2)$\\ \hline
 		OR&Returns TRUE if any argument evaluates to TRUE.&$IF(C1,V1,IF(C2,V1,IF(C3,V1,V2)))$ $\rightarrow$ $IF(OR(C1,C2,C3),V1,V2)$\\
 		\hline
 		CHOOSE&Returns a value from a list using a given position or index.&$IF(A1=1,str1,IF(A1=2,str2,IF(A1=3,str3)))$ $\rightarrow$\\&& $CHOOSE(A1,str1,str2,str3)$\\ \hline
 		MATCH&Returns a number representing a position in an array.&$IF(A1=str1,1,IF(A1=str2,2,IF(A1=str3,3)))$ $\rightarrow$\\&&
 		$MATCH(A1,\{str1,str2,str3\},0)$\\ \hline
 		LOOKUP&Perform a vertical/horizontal lookup (corresponding to function VLOOKUP and HLOOKUP) by searching for a value in the first column/row of a table and returning the value in the same row/column in the $index$ position.&$IF(A1=C1,D1,IF(A1=C2,D2,IF(A1=C3,D3,IF(A1=C4,D4))))$ $\rightarrow$ 
 		$VLOOKUP(A1,C1:D4,2,FALSE)$\\ \hline
 		MAX/MIN&Return the largest/smallest value from a supplied set of numeric values.&$IF(A>B,A,B)$$\rightarrow$$MAX(A,B)$\\ \hline
 		IFS&Run multiple tests and return a value corresponding to the first TRUE result.&$IF(C1,V1,IF(C2,V2,IF(C3,V3,IF(C4,V4))))$ $\rightarrow$ $IFS(C1,V1,C2,V2,C3,V3,C4,V4)$\\ \bottomrule	
 	\end{tabular}
 \end{table*}


In this paper, we have identified 7 categories of patterns corresponding to 7 types of spreadsheet functions. In this sub-section, we explain the patterns in details by text description, AST, and examples. The basic patterns (with \ifdepth{} of 5 in all examples) are illustrated in Figure~\ref{fig:typicalast}. Based on specific structures of each pattern, their pattern matching algorithms share the preceding general procedure and differ in minor details. 


 \textbf{(1) AND pattern.} If a \emph{nested-IF} expression satisfies the following conditions, we infer it has the semantic of the AND function, as shown in Figure~\ref{fig:typicalast}: first, the false branches of each condition are all identical; second, the true branches of each condition are all IF expressions, except for the last true value (i.e., $V1$). Such expressions can be replaced with $IF(AND(condition list), truebranch, falsebranch)$. For example, the expression with the first AST in Figure~\ref{fig:typicalast} can be replaced with $IF(AND(C1,C2,C3,C4),V1,V2)$.
%

 \textbf{(2) OR pattern.}  If a \emph{nested-IF} expression satisfies the following conditions, we infer that it actually has the semantic of the OR function, as shown in the second AST of Figure~\ref{fig:typicalast}: first, the true branches of each condition are all identical; second, the false branches of each condition are all IF expressions, except for the last false value (i.e., $V2$). Such kind of expressions can be replaced with $IF(OR(condition list), true value, false value)$. For example, the expression with the second AST in Figure~\ref{fig:typicalast} can be replaced with $IF(OR(C1,C2,C3,C4),V1,V2)$.

 \textbf{(3) CHOOSE pattern.} A \emph{nested-IF} expression that matches the semantic of CHOOSE function~\cite{choosefunction} should have the following features. First, all the conditions are number equality evaluations, with the corresponding numbers forming an arithmetic progression, which can be translated into natural sequences. Second, the false branches of each condition are all IF expressions, except for the last false value. Third, the true branch values are all strings. For example, $IF(A1=1,str1,IF(A1=2,str2,IF(A1=3,str3,IF(A1=4,str4))))$ could be transformed into $CHOOSE(A1,str1,str2,str3,str4)$; expression $IF(A1=2,str1,IF(A1=4,str2,IF(A1=6,str3,IF(A1=8,str4))))$ could be transformed into $CHOOSE(A1/2,str1,str2,str3,\\str4)$.
 

 \textbf{(4) MATCH pattern.} A \emph{nested-IF} expression that matches the semantic of MATCH function~\cite{matchfunction} should have the following features. First, all the conditions are string equality evaluations. Second, the true branch values are all numbers that could form an arithmetic progression, which can be translated into a natural sequence. Third, the false branches of each condition are all IF expressions, except for the last false value. For example, expression $IF(A1=str1,1,IF(A1=str2,2,IF(A1=str3,3,IF(A1=str4,4))))$ could be transformed into $MATCH(A1,{str1,str2,str3,str4},0)$; expression $IF(A1=str1,2,\\IF(A1=str2,4,IF(A1=str3,6,IF(A1=str4,8))))$ could be transformed into $2*MATCH(A1,{str1,str2,str3,str4},0)$.
 

\textbf{(5) LOOKUP pattern.}  A \emph{nested-IF} expression that matches the semantic of VLOOKUP/HLOOKUP \footnote{The ``V'' and ``H'' refer to ``vertical'' and ``horizontal'' respectively.} pattern~\cite{VLOOKUP,HLOOKUP} should have the following features. First, all the conditions are equality evaluations of reference values. The references are cell neighbors vertically/horizontally. Second, all the true branches are references that referred to other cells. The references are cell neighbors vertically/horizontally, and have the same columns/rows as the references in the conditions. Third, the false branches of each condition are all IF expressions, except for the last false value. For example, as shown in Table~\ref{tab:ninefunc}, expression $IF(A1=C1,D1,IF(A1=C2,D2,IF(A1=C3,D3,IF(A1=C4,D4))))$ can be transformed into $VLOOKUP(A1,C1:D4,2,FALSE)$.


The above patterns suit the circumstance that the values looked up can be found directly in other cells. For those that cannot be found directly, in this paper, we propose creating new tables in the worksheets to make ease for the look up function. Consequently, as long as the conditions are evaluating the value of a specific cell (doing look up based on this cell), we can perform transformation with the LOOKUP function. For example, for expression $IF(A1=V1,V2,IF(A1=V3,V4,IF(A1=V5,V6,IF(A1=V7,V8))))$, we create a table ranged $(E1:F4)$, where $E1=V1$, $F1=V2$, 
$E2=V3$, $F2=V4$,$E3=V5$, $F3=V6$,$E4=V7$, $F4=V8$. In this way, the expression can be transformed into $VLOOKUP(A1,E1:F4,2)$.

 \textbf{(6) MAX/MIN pattern.} A \emph{nested-IF} expression that matches the semantic of MAX or MIN pattern should have the following features. The condition should do the comparison of two parts, e.g., $A<B$, $A<=B$, $A>B$, $A>=B$. The true branch and the false branch should be these two parts respectively. For example, expressions $IF(A<B,A,B)$, $IF(A<=B,A,B)$, $IF(B>A,A,B)$, $IF(B>=A,A,B)$ can all be transformed into $MIN(A,B)$; expressions $IF(A>B,A,B)$, $IF(A>=B,A,B)$, $IF(B<A,A,B)$, $IF(B<=A,A,B)$ can all be transformed into $MAX(A,B)$.

 \textbf{(7) IFS pattern.} The IFS pattern has the fewest conditions. As long as the false branches are IF expressions (except for the leaves), the expression can be transformed with the IFS function, as shown in Table~\ref{tab:ninefunc}. Note that this pattern makes the fewest syntax changes comparing to the original syntax, and the number of conditions remain the same. However, the IFS function has the advantage of conciseness and readability, and there is also no need to worry about the IF statements and parentheses~\cite{ifs}. Additionally, there is no need to supply a value if the condition is false (unlike the \emph{nested-IF} expression which needs another IF expression to serve as the false branch)~\cite{ifstwo}. Our survey of end users reflects these advantages as well (see Section ~\ref{sec:surveyresults}).

\textbf{(8) USELESS pattern.} Except for the above patterns that match the existing spreadsheet functions, we find another pattern that does not match any function, but can also be transformed accordingly to remove nested IF. We call this pattern the \textbf{``USELESS''} pattern. For example, expression $IF(A=B,A,B)$ actually equals $A$ or $B$. We put the checking order of this patter just before the IFS pattern. 

For ease of presentation, we unify the condition redundancy, the USELESS pattern, and the 7 types of advanced functions all as ``patterns''.

\section{Evaluation}
\label{sec:evaluationdesign}

\subsection{Research Questions}
\label{sec:researchquestions}

In this paper, we investigate the following four research questions.

\noindent \textbf{RQ1:} \textbf{Are the refactored formulae functionally equal to the original ones?} This question aims to check the correctness of our refactoring approach. 


\noindent \textbf{RQ2:} \textbf{What is the refactor coverage of our approach?} This question aims to check the applicability of our approach: how many \nestifformu{} can our approach handle. 


\noindent \textbf{RQ3:} \textbf{What is the refactor effectiveness of our approach?} This question aims to check whether our refactorings relieve the nested-IF smells: how much can our approach decrease the \ifdepth{} of \nestifformu{}.


\noindent \textbf{RQ4:} \textbf{Do end users prefer the refactored formulae?} This question aims to find out the necessity of refactoring from the respective of end users, as well as whether end users prefer the refactored formula our approach provides.


\subsection{Refactor Equality}
To answer the first research question, we conduct manual inspection as well as formula calculation result comparison.

For manual inspection, considering that there are over 10 million formulae and it is impossible to check the refactorings one by one, we randomly select 2000 formula pairs $<F_o,F_r>$ ($F_o$ represents the original formula, $F_r$ represents the refactored formula). The first three authors then check each pair and record their judgements.

For formula value comparison, we scan all Excel files and replace the original \nestifformu{} with the refactored ones. For each formula pair $<F_o,F_r>$, we get a responding value pair $<V_o,V_r>$. We thus record whether $V_o$ equals  $V_r$. To automatically achieve the above process, we use $ClosedXML$, which is a powerful .NET library enabling users to create and modify Excel files.

Our experiment results indicate that either manual inspection or value comparison indicates a 100\% correctness of the refactored formulae. This result reveals the reliability of our refactoring results.

\subsection{Refactor Coverage}
\label{sec:refactorcoverage}
To answer the second research question, we present the total proportion of refactored formulae in Section~\ref{sec:totalcoverage}, the proportion of formulae handled by each pattern in Section~\ref{sec:patterncoverage}, and the proportion of formulae handled by more than one patterns in Section~\ref{sec:multicoverage}.

\subsubsection{Total Coverage}
\label{sec:totalcoverage}
For the original total set of \nestifformu{} $O_{total}$ and original unique set $O_{unique}$, we conduct automatic refactoring following the refactoring procedure introduced in Section~\ref{sec:approach}. Correspondingly, we get the refactored set $R_{total}$ (out of $O_{total}$) and $R_{unique}$ (out of $O_{unique}$). The refactor coverage can then be calculated as $\frac{\#R_{total}}{\#O_{total}}*100\%$ and $\frac{\#R_{unique}}{\#R_{unique}}*100\%$ (\# represents the number). 

Table~\ref{tab:coverage} presents the refactor coverage results. The first row is for the total formula set, the second row is for the unique formula set. From this table, our approach is able to handle almost all the \nestifformu{}, with a refactor coverage of over 99\%.

We also observe that there are around 43,000 \nestifformu{} that cannot be automatically refactored. We analyze them and found that they can be categorized into two types. In the first type, the condition part of the outmost IF expression contains another IF expression and does not match our patterns even if being treated as a whole, such as $IF(AND(IFsubexpression1,IFsubexpression2)=TRUE,value1,value2)$. In the second type, although the inner \emph{IF} expression lies in the branches of the outer expression, it is wrapped with other non-IF functions, and thus the AST is quite complex, such as $IF(Condition, SUM(IFsubexpression1, IFsubexpression2), value)$.

\begin{table}\small
	\centering
	\caption{\label{tab:coverage} Refactor coverage}
	\begin{tabular}{l|l|l|l} 
		\toprule
		\textbf{Formula Set}&\textbf{Original}&\textbf{Refactored}&\textbf{Refactor Coverage}\\
		\midrule
		Total&27,689,299&27,645,688&99.84\% \\ 
		Unique&19,260,407&19,243,407&99.91\% \\ \bottomrule	
	\end{tabular}
\vspace{-4mm}
\end{table}

\subsubsection{Coverage of Each Single Pattern}
\label{sec:patterncoverage}
We next investigate the coverage of each pattern. To do this, for each refactored formula, we assign it with a pattern list $pList=[REDUN, AND, OR, CHOOSE,\\ MATCH, LOOKUP, MAXMIN, USELESS, IFS]$ ($REDUN$ represents redundancy) recording which patterns are adopted during the refactoring process. Because the $MAX$ and $MIN$ are quite similar, we merge them into one $MAXMIN$. The adopted patterns are assigned with a value of 1, otherwise their values are 0. 

For example, for formula $IF(condition1, value1, IF(cell1>cell2,\\cell1,cell2))$, our approach will refactor it into $IF(condition1,value1,\\MAX(cell1,cell2))$,and thus $pList = [0,0,0,0,0,0,1,0,0]$. Each formula would have a list. In this way,
we will get a matrix of all patterns with the total formula set, based on which it is easy to calculate the proportion of the formulae that each pattern handles. 

We present the final results in Table~\ref{tab:patternproportion}. Column ``Total'' shows the results of the total formula set. Column ``Unique'' shows the results of the unique formula set. Column ``Coverage'' shows the proportion of formulae that each pattern handles against all original formulae. The first row presents the results of redundant conditions, the remaining rows show the results of different alternative functions.

As shown in Table~\ref{tab:patternproportion}, around 6\% formulae contain redundant conditions. This is rather surprising, because redundant conditions are somewhat low-level spreadsheet smells users should not make if they know the basic structure of \emph{IF} expressions. The dead branches caused by redundant conditions are like the dead code in traditional programming, and should be removed definitely. These results reflect the fact that end users may lack the basic knowledge of programming, even in understanding conditional logic.

For the patterns, the AND, OR, and LOOKUP patterns handle around 6\%-15\% of formulae respectively, the IFS pattern handles as high as 65\% of formulae. The remaining patterns such as MATCH, CHOOSE, and USELESS handle less than 2\% respectively. The high refactor coverage of the IFS pattern is not surprising, because as we introduced in Figure~\ref{fig:typicalast}, IFS has the fewest conditions which most \nestifexp{} match. 

\begin{table}\small
	\centering
	\caption{\label{tab:patternproportion} Refactor coverage of each pattern}
	\begin{tabular}{l|rr|rr} 
		\toprule
		\multirow{2}{*}{\textbf{Pattern}}&\multicolumn{2}{c}{\textbf{Total}}&\multicolumn{2}{|c}{\textbf{Unique}} \\ \cline{2-5}
		&\textbf{Refactored}&\textbf{Coverage}&\textbf{Refactored}&\textbf{Coverage}\\ \midrule

		REDUN&1,766,803&6.39\%&1,250,006&6.50\%\\ \hline
		AND&2,322,346&8.40\%&2,206,296&11.47\%\\
		OR&4,169,992&15.08\%&2,099,450&10.91\%\\
		CHOOSE&165,594&0.60\%&141,695&0.74\%\\
		MATCH&23,254&0.08\%&9,331&0.05\%\\
		LOOKUP&1,780,419&6.44\%&1,637,564&8.51\%\\
		MAXMIN&234,960&0.85\%&214,452&1.11\%\\
		USELESS&83,060&0.30\%&69,912&0.36\%\\
		IFS&18,239,046&65.97\%&12,520,577&65.06\% \\
		\bottomrule	
		
	\end{tabular}
\vspace{-2mm}
\end{table}

\subsubsection{Coverage of Multi Patterns}
\label{sec:multicoverage}

As our approach will repeatedly try different patterns until no \emph{IF} expressions can be removed, some formulae may be handled by more than one pattern during the repeat. Corresponding to the $pList$ (introduced in Section~\ref{sec:patterncoverage}) of this formula, more than one element would have value ``1''. To investigate the frequency of such circumstances, we record the formulae handled by multi patterns. 

\begin{table}\small
	\caption{\label{tab:multipattern}Number of formulae processed by multi-patterns}
	\begin{tabular}{p{3cm}|r|r}
		\toprule
		\textbf{Pattern Num}&\textbf{Total}&\textbf{Unique} \\ \midrule
		2& 1,019,035~(3.69\%)&785,156~(4.1\%)\\ 
		3&60,843~(0.22\%)&60,825~(0.32\%)\\ \hline 
		total&1,079,878 ~(3.91\%) & 845,981~ (4.40\%)\\ \bottomrule
	\end{tabular}
\vspace{-2mm}
\end{table}

The results are shown in Table~\ref{tab:multipattern}. In summary, at most 3 patterns are applied to process a formula. 3.91\%/4.40\% of the total/unique set of refactored formulae are processed with more than one patterns. We also present the specific overlap of different patterns (of the total set), as shown in Figure~\ref{fig:circos}. The figure is a circos\footnote{\url{http://mkweb.bcgsc.ca/tableviewer/visualize/}} visualization. The pattern segment size reflects the scale of refactored formulae of each pattern (shown in Table~\ref{tab:coverage}). The lines between different segments are called cells. The thickness of these cells can indicate the absolute number of overlapped formulae between different patterns. As shown in this figure, the IFS pattern has overlap with almost all other patterns, except AND, MATCH, CHOOSE. This is because the IFS pattern has the fewest conditions and many formulae satisfy them (as the example in Section~\ref{sec:premilinarystudy} shows). MATCH has no overlap with any pattern. The reason may be that formulae that match the MATCH pattern are usually very regular and have simple semantics.

\begin{figure}[t]
	
	\includegraphics[scale=0.3]{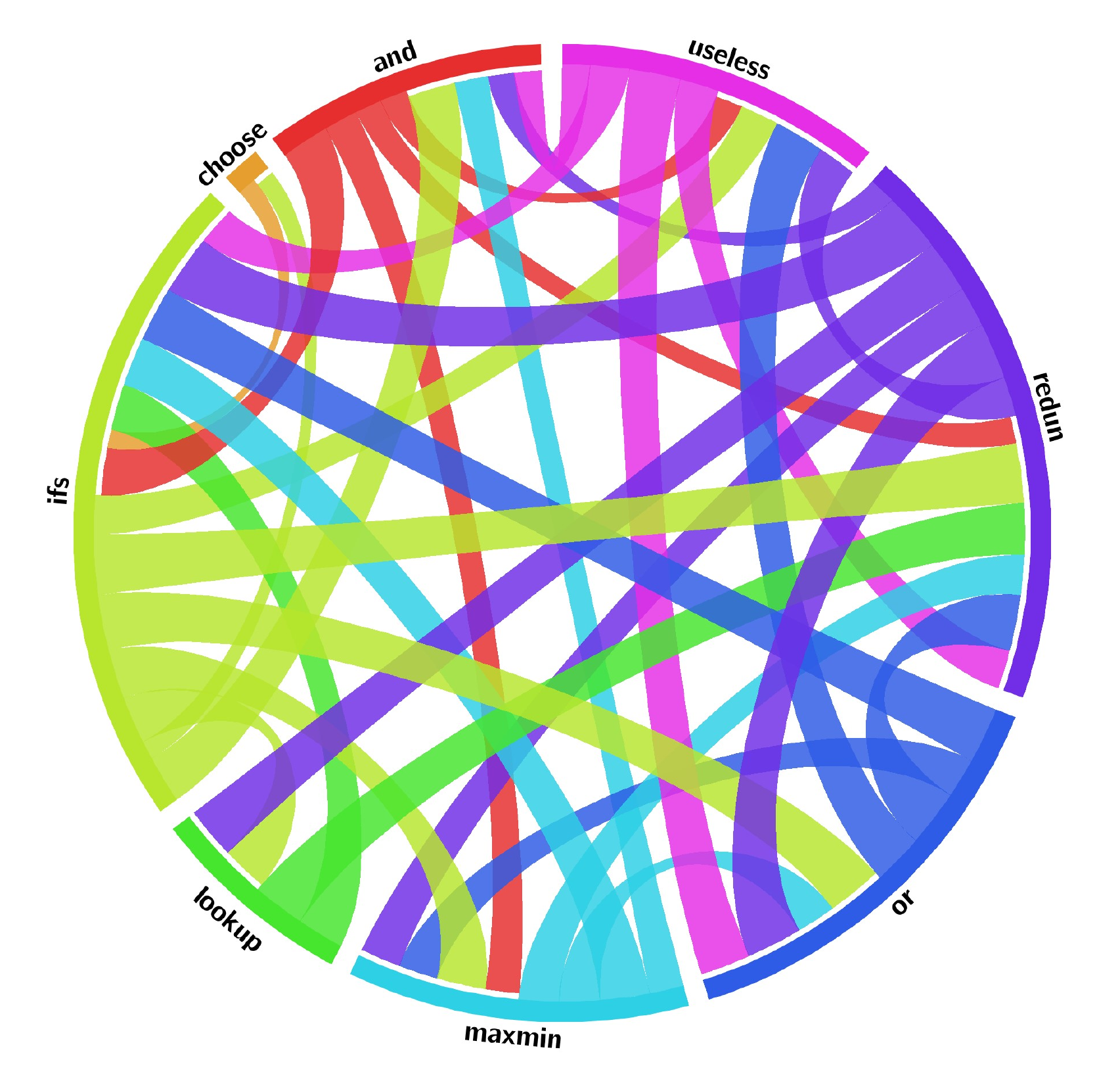}
	\caption{\label{fig:circos}Circos chart of the overlap between patterns.}
	\vspace{-2mm}
\end{figure}
%
%
%

\subsection{Refactor Effectiveness}
\label{sec:effectiveness}

To answer the third research question, we present the absolute \ifdepth{} reduction results in Section~\ref{sec:absolutedepthreduce}, the relative \ifdepth{} reduction (the depth reduction rate) in Section~\ref{sec:relativedepthreduce}, and the final \ifdepth{} of refactored formulae in Section~\ref{sec:finaldepth}.

\subsubsection{Absolute Depth Reduction}
\label{sec:absolutedepthreduce}
First, we would like to know how many \emph{if-depths} can our approach reduce on the refactored formulae. For each refactored formula pair $<F_o,F_r>$, we parse $F_o$ and $F_r$ and calculate their respective \ifdepth{}: $dep_o$, $dep_r$. The depth reduction $DepReduce_{num}$ is then calculated by $DepReduce_{num}=dep_o-dep_r$. 

\begin{table*}[h]\small
	\centering
	\caption{\label{tab:absolutedepthreduce} Absolute number of \ifdepth{} reduction for Unique.}
	\begin{tabular}{p{1.6cm}|r|r|r|r|r|r|r|r|r|r|r|r|r} \hline
		Depth Reduce&1&2&3&4&5&6&7&8&9&10&11&12&13\\ \cline{1-14}
 Number&7,885,404&5,284,297&2,280,668&836,304&418,525&96,353&1,790,044&454,225&38,614&5,772&45,452&47,105&6,927\\ \cline{1-14}
Depth Reduce&14&15&16&19&20&22&24&27&28&29&36&39&48\\ \cline{1-14}
 Number&6,302&12,214&507&10,200&1&12&10,200&1,963&490&490&2&795&10,541\\ \hline
		
	\end{tabular}
\vspace{-2mm}
\end{table*}

Table~\ref{tab:absolutedepthreduce} presents the results for the unique set. Row ``Formula Number'' presents the number of formulae that was refactored with the corresponding depth reduction. For example, in the first cell, the number 7,885,404 means that among all the refactored formulae, 7,885,404 of them have a \ifdepth{} reduction of 1. From the table, our approach is able to reduce the \ifdepth{} with various degrees, indicating the effectiveness of our approach.

Most of the refactorings have a depth reduction of below 5. However, the absolute depth reduction results shown in this table depend on the amount of original formulae with different \emph{if-depths}. A formula with 2 \nestiffunc{} can have a depth reduction of 2 at most. To relieve this problem, we also present the relative depth reduction results in Section~\ref{sec:relativedepthreduce} as well as the final new \ifdepth{} of the refactored formulae in Section~\ref{sec:finaldepth}.

\subsubsection{Relative Depth Reduce}
\label{sec:relativedepthreduce}
As mentioned above, only presenting the absolute number of depth reduction may fail to reflect the real effectiveness. In this section, we also present the results of relative depth reduction: $DepReduce_{ratio}=DepReduce_{num}/dep_o$. For ease of presentation, we divide $DepReduce_{ratio}$ into four ranges: $(0\%,25\%]$ \footnote{0\% $<DepReduce_{ratio}<=$ 25\%}, $(25\%,50\%]$, $(50\%,75\%]$, and $(75\%,100\%]$. The distribution of each range is presented in Figure~\ref{fig:depthreduceprop}.

From the figure, in general over half of the refactoring reduction is over 50\% of \ifdepth{}. In particular, for the total and unique set of refactored formulae, most refactorings fall into the range of $(25\%,50\%]$ and $(75\%,100\%]$, indicating that our relative depth reduction results are good.
\newcommand{\scaa}{0.21}
\begin{figure}[!tb]
	\center	
	\begin{tabular}{cc}
		
		\hspace*{-8mm}\includegraphics[scale=\scaa]{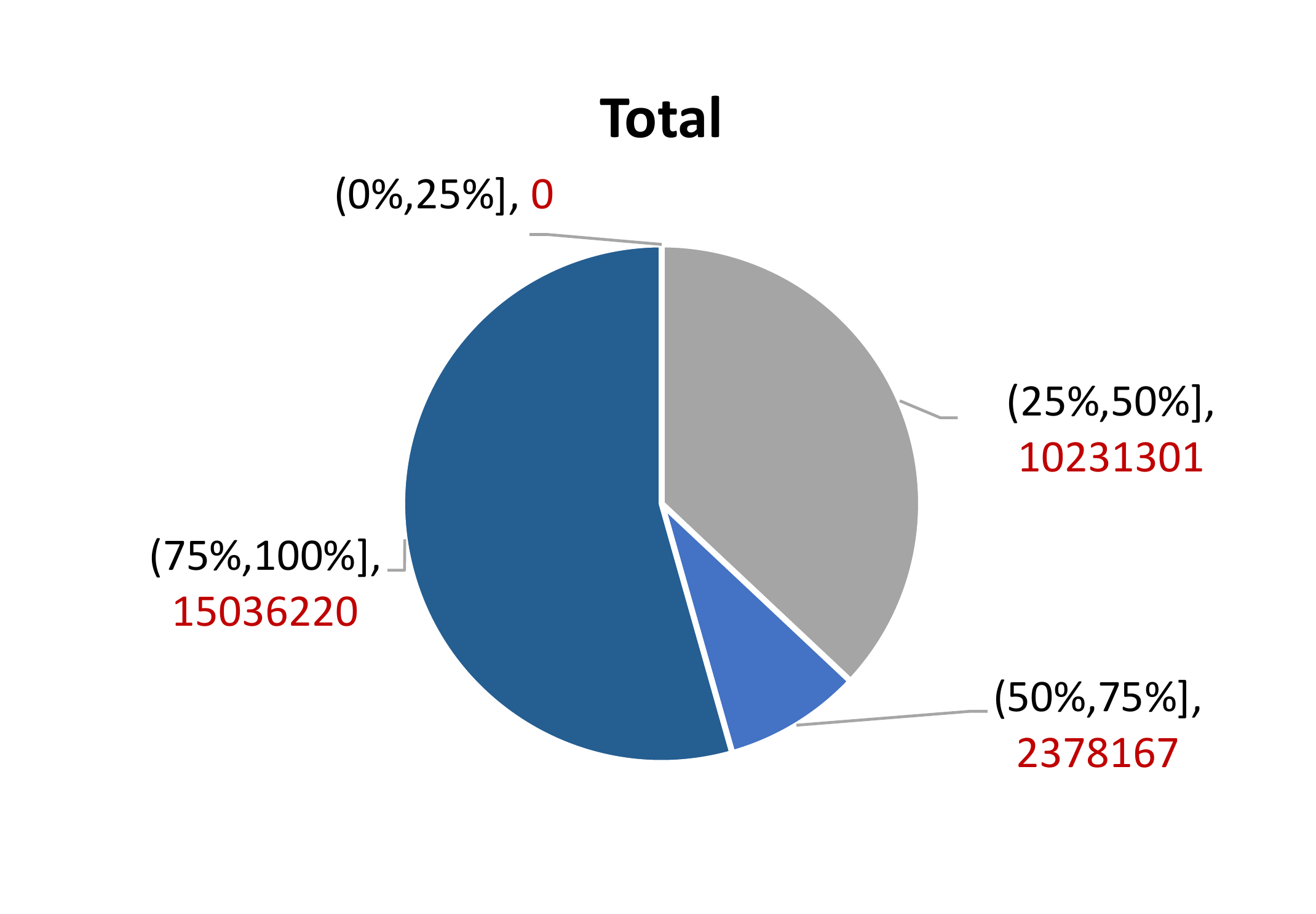}&
		\hspace*{-5mm}\includegraphics[scale=\scaa]{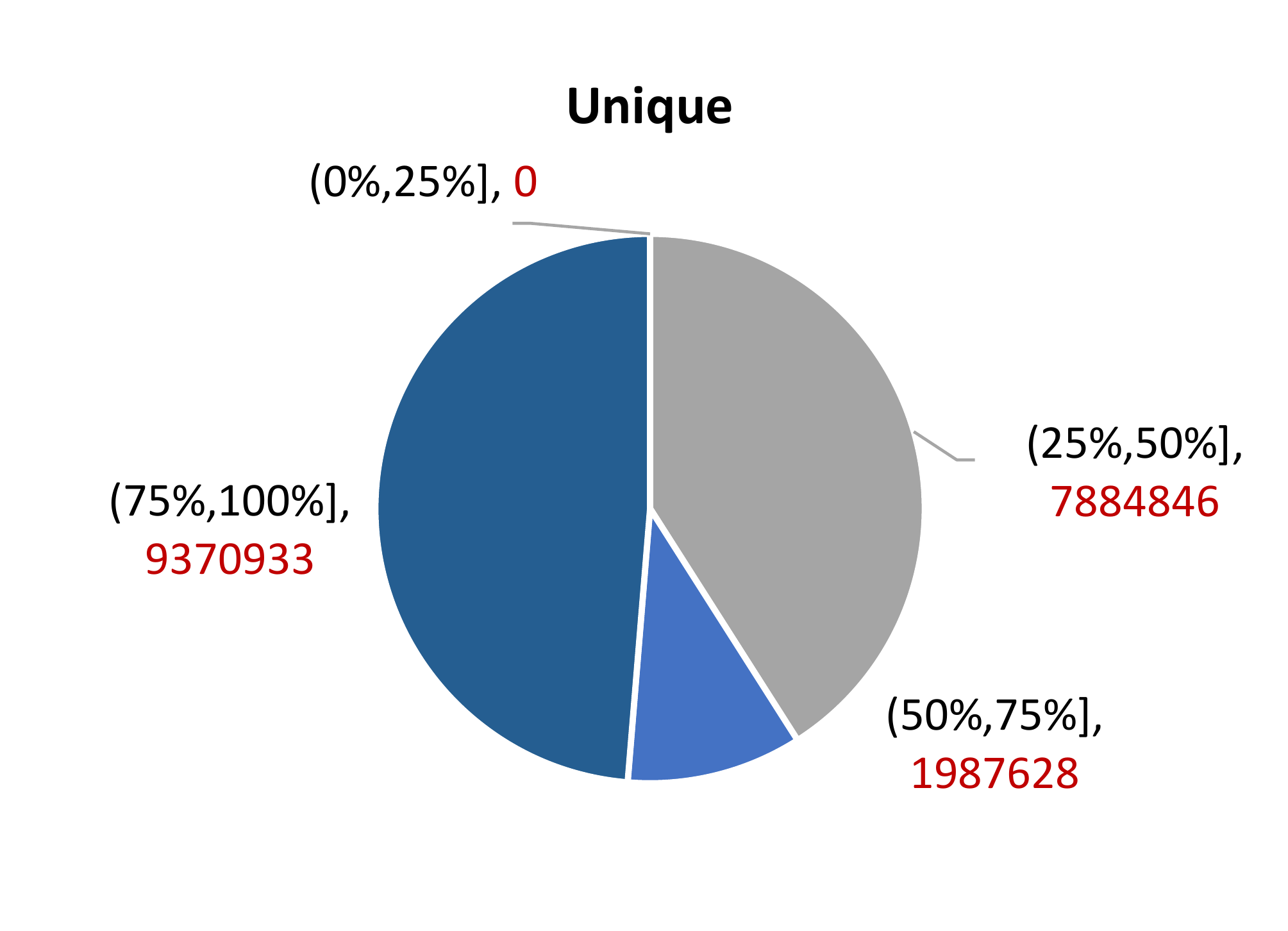}\\
	\end{tabular}
	\caption{\label{fig:depthreduceprop} Distribution of different depth reduction ratios. }	
	\vspace{-2mm}
\end{figure}

\subsubsection{Final Depth After Refactoring}
\label{sec:finaldepth}
Except for the absolute and relative depth reduction results, we check whether the refactored formulae still have large \ifdepth{}. To answer this question, we investigate the new \ifdepth{} $dep_r$ of each refactored formula $F_r$. The results are shown in Table~\ref{tab:finaldepth}.

From the table, most of the refactorings yield a new \ifdepth{} of 0 or 1\footnote{\ifdepth{} of 0 and 1 are equally effective in relieving nested IF smells, because either of them avoid the smell completely.}, indicating that our approach is able to completely remove the \nestiffunc{} in most formulae. 


\begin{table}[t!]\small
	\centering
	\caption{\label{tab:finaldepth} Number of formulae with different new depth}
	\begin{tabular}{r|r|r|r} 
		\toprule
		\multicolumn{2}{c}{\textbf{Total}}&\multicolumn{2}{|c}{\textbf{Unique}} \\ \hline
		New Depth & Formula Num & New Depth  & Formula Num \\  \hline
		
		0&13,906,460~(50.30\%)&0&8,723,082~(45.33\%) \\ 
		1&	13,717,158~(49.62\%)	&1&	10,498,258~(54.56\%)\\ 
		2&	22,070&	2&	22,067\\ 
	 \bottomrule
	\end{tabular}
	\vspace{-2mm}
\end{table}

\subsection{Preference of End Users}
\label{sec:surveyresults}
In this section, we explore end users' attitude towards the \nestifformu{} and the refactored formulae. Our survey includes 49 participants. 26 of them have been using use spreadsheets quite often (over 5 times per-week), 18 of them use spreadsheet sometimes (about once per-week), 2 occasionally (about once per-month), and 3 rarely use them. Additionally, about two thirds of them are employees employed in banks, law firms, telecommunication companies, and so on. The remaining ones are mainly college students.

Our survey includes seven parts. Each part contains one representative pattern\footnote{Pattern AND and OR are combined, as are MAX and MIN.}. For each pattern, we present two functionally equivalent spreadsheet formulae $F1$ and $F2$ and several questions concerning the participants' preferences.
For example, for the AND pattern, we first describe a function like ``If three conditions Condition1, Condition2, Condition3 are all TRUE, return Value1; otherwise return Value2.''. Then, $F1$ and $F2$ will be presented as: $
F1: IF(Condition1,IF(Condition2,IF(Condition3,Value1,Value2),\\Value2),Value2)
$, $F2: 
IF(AND(Condition1, Condition2, Condition3), \\Value1, Value2)
$. At the beginning, the participants are invited to have a look at a formula pair. Then, they are supposed to answer four questions as shown in the first column of Figure~\ref{fig:survey}. Question Q1 investigates participant's basic knowledge. Q2 investigates participant's preference (between $F_o$ and $F_r$). Q3 and Q4 checks whether a participant lacks the knowledge of manual refactoring and whether automatic refactoring is needed. As each of the 49 participants are supposed to finish the seven parts one by one, for each question we collect $49*7 =343$ answers. We call each answer a ``case''.

\begin{table*}\small
	\centering
	\caption{\label{tab:surveyresults} Feedback from end users}
	\vspace{-1mm}
	\begin{tabular}{p{5.5cm}|r|r|r|r|r|r|r||r|l} 
		\toprule
		Answer&Redundancy&AND/OR&CHOOSE&MATCH&LOOKUP&MAX/MIN&IFS&\textbf{Total}&\textbf{Prop.}\\ \midrule 
		A1: I know F1 equals to F2.&17&18&12&10&8&24&9&98&28.57\%
		\\ \hline
		A2: I prefer F2.&45&46&46&48&43&45&45&318&92.71\% \\ 
		A2.A:F2 is shorter.&24&23&26&26&28&27&23&177&51.60\%
		\\ 
		A2.B:F2 is less complex.&41&41&29&29&34&31&37&242& 70.55\%
		\\ 
		A2.C:F2 is easier to understand.&31&26&29&29&24&27&24&190&55.39\%
		\\ 
		A2.D:F2 is not easy to make mistakes.&17&18&20&20&33&22&17&147&42.86\%
		\\ \hline
		A3: I can refactor manually.&11&11&10&9&7&20&4&72 &20.99\%
		\\ \hline
		A4: Automatic refactoring is helpful. &48&49&48&48&49&49&48&339&98.83\%
		\\ \bottomrule
	\end{tabular}
	\vspace{-3mm}
\end{table*}

\begin{figure}
	\includegraphics[scale=0.42]{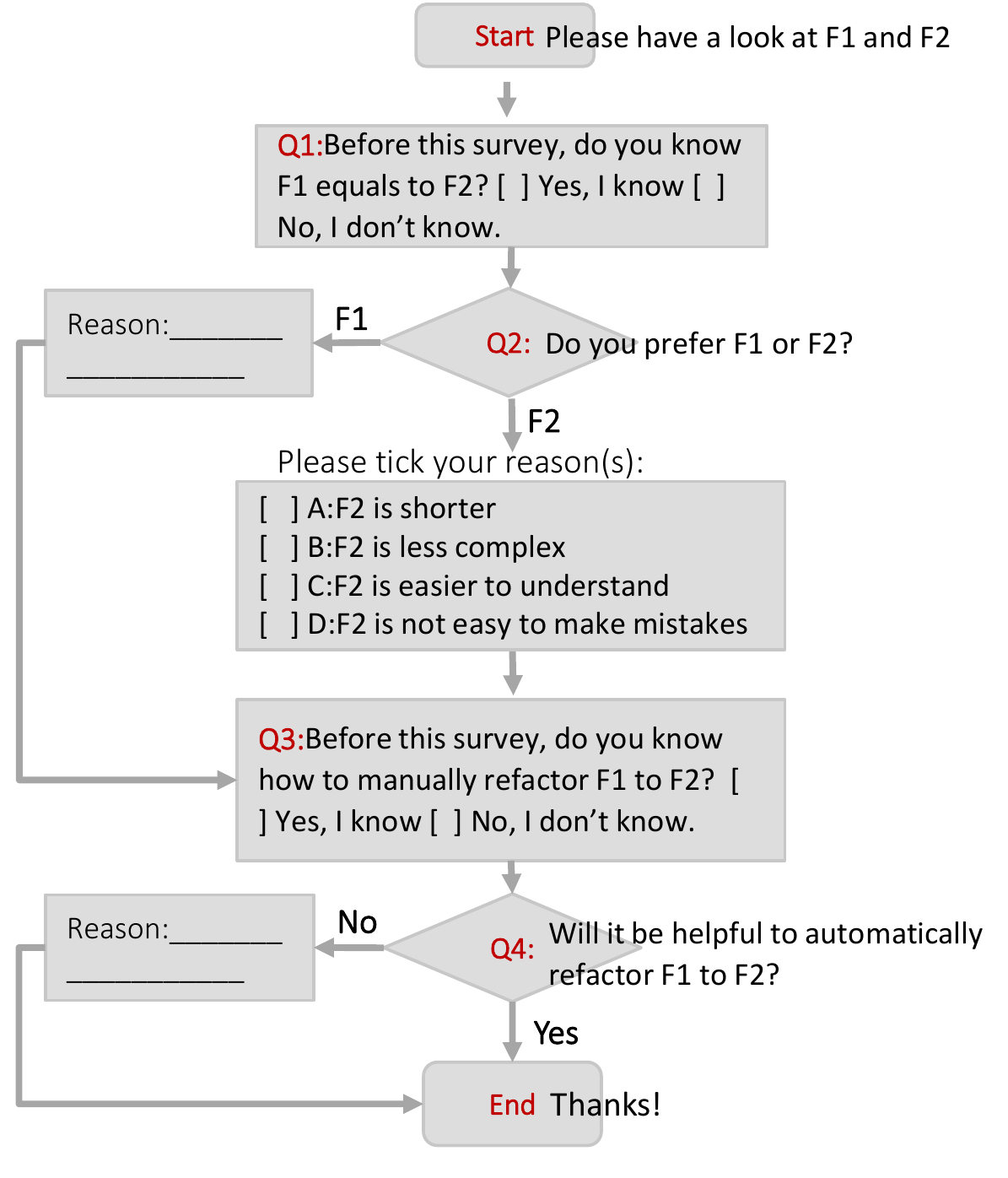}
	\caption{\label{fig:survey} Survey questions.}
	\vspace{-5mm}
\end{figure}

The survey results are shown in Table~\ref{tab:surveyresults}. Row ``A2.A''~``A2.D'' are concerned with the reasons why participants prefer $F2$ (the refactored formula). We first focus on the total survey results shown in Column ``Total'' and ``Prop.''. From this table, only under 28.57\% cases participants have the knowledge to judge the equivalence between $F1$ and $F2$. 92.71\% of cases participants prefer the refactored formulae. All four reasons we listed have high votes, with ``A2.B:F2 is less complex.'' the highest. Only with 20.99\% cases participants have the ability to manually refactor. 98.83\%  believe that our automated refactoring approach is necessary and helpful.

Note that there are still around 7\% (24) of the cases where participants do not like the refactored formulae. We look into their comments and find that in 6 cases the participants prefer $F1$ because they have no idea which one is better and choose one answer randomly. In one case the participant said that $IF(cell1>cell2,cell1,cell2)$ is better than $MAX(cell1,cell2)$ because when cell1 equals cell2, the result is more specific. In all the remaining 17 cases the participants prefer $F_1$ because they do not have knowledge related to $F_2$. Similarly, there are 1\% cases where participants regard our approach helpless, and all of the reasons are that they lack knowledge about the refactored formulae. These negative opinions are quite valuable, indicating that in practical application, it is necessary to provide knowledge related to new functions to help end users understand them better. A possible application scenario may be that refactoring will not be conducted without the permission of end users. They may get some refactor suggestions as well as explanations, and could choose whether they would like to adopt the suggestion. More discussion about the application scenario can be found in Section~\ref{sec:discussion}.

We then focus on the answers of different patterns. LOOKUP and IFS are known to the fewest participants, while MAXMIN is known to the most participants. More people think that LOOKUP are less error-prone comparing to other patterns. The other answers more or less have no obvious differences among different patterns.

\section{Discussion}
\label{sec:discussion}
In this section, we discuss the application scenario (in Section~\ref{sec:applicationscenario}) and the future work of our approach (in Section~\ref{sec:futurework}).
%
%
%

\subsection{Application}
\label{sec:applicationscenario}

Our application scenario is to make our approach a spreadsheet plug-in. When an end user finishes writing a formula with nested IF functions, the plug-in may identify whether the formula can be refactored. If yes, it alerts that these nested IFs are bad smells, and provides refactor suggestions. Note that in this paper each formula will yield one specific refactor result, while it is also applicable that we generate different refactoring results: for the formulae that can be handled by multiple patterns, each pattern corresponds to one result. All the results can be ranked and serve as suggestion candidates, so that the end users can choose which ever they like. Note that from our survey results introduced in Section~\ref{sec:surveyresults}, it maybe better if we provide explanations of each suggestion to aid the understanding of end users. We will develop and present such a plug-in in future work.

\subsection{Limitations and Future Work}
\label{sec:futurework}
	
There are several directions that our approach can be improved. 

First, there may be better application scenarios. Except for the application scenario introduced above, another more intelligent one is to provide possible refactoring suggestions before the end users finish writing the formula. This scenario especially suits (intended) \nestifformu{} with high \ifdepth{}. For example, suppose that an end user intends to write a 30-depth Nest-IF formula, it would be useful if we identify the possible semantics when the user starts to write the 10th nested IF expression. In this way, the user may skip the verbose writing of the remaining 20 depth and choose to use the suggested function. Another advantage is that this application scenario can help users to avoid making errors aroused from using \nestifformu{}. This application scenario is applicable. In future we plan to use machine learning techniques to infer the semantics of a formula based on a part of its syntax.

Second, the refactor effectiveness can be further improved. Although our approach is currently effective in relieving the smell of \nestifformu{}, the coverage of some patterns can be enlarged. For example, for the formula in Section~\ref{sec:premilinarystudy}: $IF(Q1=X1,Q1,IF(Q1=``",X1,Q1)$, except for the IFS pattern, it would also match the OR pattern if we transform it into $IF(Q1=X1,Q1,IF(Q1!=``",Q1,X1)$ (refactored as $IF(OR(Q1=X1,Q1!=``"),Q1,X1)$). In other words, it is interesting to explore how to preprocess a formula to make it better prepared for refactoring. 


Third, the refactor coverage can be further improved. There are some \nestifformu{} out of our refactoring ability, such as the two types mentioned in Section~\ref{sec:refactorcoverage}. Also, our current patterns are not the universal set. There must be some other patterns that can also contribute to the nested IF problem, yet are outside of our current knowledge. Nevertheless, the main idea of the approach remains effective, and we would complement the pattern set upon finding other good pattern candidates.

%
%

\section{Related Work}
\label{sec:relatedwork}
End user programming has been studied since 1993~\cite{nardi1993small}. Spreadsheets have been recognized as the most successful and most popular form of end user programming~\cite{7476773}. Current research on spreadsheets mainly follows the trend of applying traditional software engineering methods to deal with spreadsheet problems. For example, most research focus on smell detection~\cite{abreu2014smelling,6227171,6405300,Hermans2015,cunha2012towards,fisher2005euses,dou2014spreadsheet,cheung2016custodes,6976077}, fault detection and automatic repair~\cite{7884634,panko2016we,rajalingham2008classification,ayalew2008spreadsheet,abreu2014faultysheet,ruthruff2006interactive}, clone detection~\cite{Hermans:2013:DCD:2486788.2486827,hermans2013data,roy2009detection}, refactoring~\cite{Hermans2015,dou2014spreadsheet,hermans2014bumblebee},  visualisation~\cite{Hermans:2011:SPS:1985793.1985855,hermans2010automatically,igarashi1998fluid,shiozawa19993d}, and so on.

The research work most related to ours include smell detection and refactoring. The former is related to the motivation of this paper: why \nestifformu{} are bad smells. The latter is related to the approach of this paper: how to refactor spreadsheets to reduce smells. We next introduce these two aspects one by one.

\subsection{Smell Detection}
\label{sec:smelldetection}
Same as code smells~\cite{fowler1999refactoring}, spreadsheets smells refer to some characteristics that may cause problems. Smells have different levels: formula-level, cell level, and structural level. We mainly introduce the formula-level ones. 

Abreu et.al.~\cite{6976077} combines 15 smells to indicate potential faults. They treat conditional complexity as one of the key smells. The results indicate that this smell only can detect 6 spreadsheet faults. 

Hermans et.al.~\cite{Hermans2015} regard conditional complexity as one of the five smells, because even in traditional professional programming, conditional complexity is a threat to code readability. However, according to their results derived from EUSES, on average each spreadsheet only has 3 formulae containing at least one condition, while from our corpus, we find that on average each spreadsheet has 1,193 formulae containing conditions; from the corpus of Enron, the number is 217. The reason for this huge difference may be that EUSES contains a lot of toy spreadsheets created by users who rarely use spreadsheet formulae.

Hermans et.al.~\cite{Hermans2015} also mention that end users already know the bad effects of conditional complexity. Our survey results  confirm this statement: around half of the participants think that formulae with high conditional complexity are more complex and error-prone; 70.55\% think that they are harder to understand.

Another work of Hermans et.al.~\cite{7476773} present an overview of software engineering approaches applied to spreadsheets. They claim that most spreadsheets contain formulae with multiple IF conditions, which is an obvious spreadsheet smell. 

\subsection{Formula Refactoring}
\label{sec:formularefactoring}

Badame and Dig~\cite{badame2012refactoring} are the first to propose refactoring in the spreadsheet domain. A tool -- ReeBook --  is presented, with which seven refactoring patterns are presented. These seven patterns target at different smells. For example, pattern \emph{MAKE CELL CONSTANT} aims to make formulae less error prone and more readable by adding the $\$$ symbol. However, their approach is disperse and can handle only simple formulae. For example, one of their refactoring patterns is called ``REPLACE AWKWARD FORMULA'', which only focus on the SUM function (e.g., replace $B5+C5+D5+E5$ with $SUM(B5:E5)$ ). They evaluate their approach on EUSES corpus and find that their refactoring can be applied to many formulae. However, they only present the number of formulae that are ``potential candidates'' for each pattern, while not presenting the actual number of successfully refactored formulae. Thus, the refactor coverage and effectiveness are unknown. 

Hermans et.al.~\cite{Hermans2015} defined different refactoring according to their smells. The results indicate that their refactoring approach is able to relieve the smells of 87\% formulae. However, their approach does not support automated refactoring.

Later on, Hermans and Dig~\cite{hermans2014bumblebee} combine the two approaches above and present BumbleBee, which is a refactoring tool allowing a formula to be refactored based on the defined transformation rules. Several patterns such as MAXMIN and OR are also mentioned in the paper. However, the formula can be refactored only when the transformation rule is defined, while according to our survey, only 20.99\% of participants may have the knowledge of defining transformation rules. The work of Hoepelman~\cite{hoepelman2015tool} expand this work and introduces more refactoring support.

To sum up, currently several works aim to tackle the challenges brought by spreadsheet smells, while no automatic and high-coverage refactoring approach is available. We propose to systematically tackling the \nestifformu{} refactoring problem, which is able to handle almost all formulae with high depth-reduce effectiveness. 

\section{Conclusion}
\label{sec:conclusion}

We propose a spreadsheet formula refactoring approach aiming to automatically relieve the smells of nested IF functions. We first try to identify if the formula contains redundant conditions. Afterwards, we identify the semantics of the combination of nested IFs and replace them with an alternative spreadsheet function. Evaluation on a very large real world spreadsheet corpus indicates that the refactor effectiveness is impressive: most of the \nestifformu{} can be refactored. Our survey of 49 participants reveals that the majority of them like our refactoring approach and think it is helpful.  

\newpage
\balance
\bibliographystyle{unsrt}
\bibliography{sample-bibliography} 

\begin{thebibliography}{10}

\bibitem{spreadmostpopular}
wiki.
\newblock {End user development}.
\newblock \url{https://en.wikipedia.org/wiki/End-user_development}, 2015.

\bibitem{burnett10}
Margaret~M Burnett and Christopher Scaffidi.
\newblock 10. end-user development.

\bibitem{hermans2015detecting}
Felienne Hermans, Martin Pinzger, and Arie van Deursen.
\newblock Detecting and refactoring code smells in spreadsheet formulas.
\newblock {\em Empirical Software Engineering}, 20(2):549--575, 2015.

\bibitem{Hermans2015}
Felienne Hermans, Martin Pinzger, and Arie van Deursen.
\newblock Detecting and refactoring code smells in spreadsheet formulas.
\newblock {\em Empirical Software Engineering}, 20(2):549--575, Apr 2015.

\bibitem{6976077}
R.~Abreu, J.~Cunha, J.~P. Fernandes, P.~Martins, A.~Perez, and J.~Saraiva.
\newblock Smelling faults in spreadsheets.
\newblock In {\em Proc. ICSME}, pages 111--120, Sept 2014.

\bibitem{7476773}
F.~Hermans, B.~Jansen, S.~Roy, E.~Aivaloglou, A.~Swidan, and D.~Hoepelman.
\newblock Spreadsheets are code: An overview of software engineering approaches
  applied to spreadsheets.
\newblock In {\em Proc. SANER}, volume~5, pages 56--65, March 2016.

\bibitem{Tufano:2015}
Michele Tufano, Fabio Palomba, Gabriele Bavota, Rocco Oliveto, Massimiliano
  Di~Penta, Andrea De~Lucia, and Denys Poshyvanyk.
\newblock When and why your code starts to smell bad.
\newblock In {\em Proc. ICSE}, pages 403--414. IEEE Press, 2015.

\bibitem{19_tips_for_nested_if_formulas}
Dave Bruns.
\newblock {19 tips for nested IF formulas}.
\newblock \url{http://https://exceljet.net/nested-ifs}, 2016.

\bibitem{reddit2015}
reddit.
\newblock {Is it a good or bad practice reducing nested if statements}.
\newblock
  \url{https://www.reddit.com/r/csharp/comments/33puzj/is_it_a_good_or_bad_practice_reducing_nested_if/},
  2015.

\bibitem{reddit_neverusenf}
reddit.
\newblock {Never use nested IFs again}.
\newblock
  \url{https://www.reddit.com/r/excel/comments/2slys1/never_use_nested_ifs_again/},
  2015.

\bibitem{hermans2015enron}
Felienne Hermans and Emerson Murphy-Hill.
\newblock Enron's spreadsheets and related emails: A dataset and analysis.
\newblock In {\em Proc. ICSE}, pages 7--16. IEEE Press, 2015.

\bibitem{badame2012refactoring}
Sandro Badame and Danny Dig.
\newblock Refactoring meets spreadsheet formulas.
\newblock In {\em Proc. ICSM}, pages 399--409. IEEE, 2012.

\bibitem{hermans2014bumblebee}
Felienne Hermans and Danny Dig.
\newblock Bumblebee: a refactoring environment for spreadsheet formulas.
\newblock In {\em Proc. ICSE}, pages 747--750. ACM, 2014.

\bibitem{fisher2005euses}
Marc Fisher and Gregg Rothermel.
\newblock The euses spreadsheet corpus: a shared resource for supporting
  experimentation with spreadsheet dependability mechanisms.
\newblock In {\em ACM SIGSOFT Software Engineering Notes}, volume~30, pages
  1--5. ACM, 2005.

\bibitem{aurigemma2010detection}
Salvatore Aurigemma and Raymond~R Panko.
\newblock The detection of human spreadsheet errors by humans versus inspection
  (auditing) software.
\newblock {\em arXiv preprint arXiv:1009.2785}, 2010.

\bibitem{gazoni2016openpyxl}
E~Gazoni and C~Clark.
\newblock openpyxl-a python library to read/write excel 2010 xlsx/xlsm files,
  2016.

\bibitem{7739679}
T.~Schmitz and D.~Jannach.
\newblock Finding errors in the enron spreadsheet corpus.
\newblock In {\em Proc. VL/HCC}, pages 157--161, 2016.

\bibitem{jansen2015enron}
Bas Jansen.
\newblock Enron versus euses: A comparison of two spreadsheet corpora.
\newblock {\em arXiv preprint arXiv:1503.04055}, 2015.

\bibitem{reschenhofer2017conceptual}
Thomas Reschenhofer, Bernhard Waltl, Klym Shumaiev, and Florian Matthes.
\newblock A conceptual model for measuring the complexity of spreadsheets.
\newblock {\em arXiv preprint arXiv:1704.01147}, 2017.

\bibitem{choosefunction}
CHOOSE.
\newblock {CHOOSE function}.
\newblock
  \url{https://support.office.com/en-us/article/CHOOSE-function-fc5c184f-cb62-4ec7-a46e-38653b98f5bc}.

\bibitem{matchfunction}
MATCH.
\newblock {MATCH function}.
\newblock
  \url{https://support.office.com/en-us/article/MATCH-function-e8dffd45-c762-47d6-bf89-533f4a37673a}.

\bibitem{VLOOKUP}
VLOOKUP.
\newblock {VLOOKUP function}.
\newblock
  \url{https://support.office.com/en-us/article/VLOOKUP-function-0bbc8083-26fe-4963-8ab8-93a18ad188a1}.

\bibitem{HLOOKUP}
HLOOKUP.
\newblock {HLOOKUP function}.
\newblock
  \url{https://support.office.com/en-us/article/HLOOKUP-function-a3034eec-b719-4ba3-bb65-e1ad662ed95f}.

\bibitem{ifs}
office.
\newblock {IF function – nested formulas and avoiding pitfalls }.
\newblock
  \url{https://support.office.com/en-us/article/IF-function-%e2%80%93-nested-formulas-and-avoiding-pitfalls-0b22ff44-f149-44ba-aeb5-4ef99da241c8?ui=en-US&rs=en-US&ad=US},
  2015.

\bibitem{ifstwo}
spreadsheeto.
\newblock {Let’s take a look at IF, Nested IF and IFS}.
\newblock \url{http://spreadsheeto.com/if/}, 2015.

\bibitem{nardi1993small}
Bonnie~A Nardi.
\newblock A small matter of programming: perspectives on end user programming,
  1993.

\bibitem{abreu2014smelling}
Rui Abreu, J{\'a}come Cunha, Joao~Paulo Fernandes, Pedro Martins, Alexandre
  Perez, and Jo{\~a}o Saraiva.
\newblock Smelling faults in spreadsheets.
\newblock In {\em Proc. ICSME}, pages 111--120. IEEE, 2014.

\bibitem{6227171}
F.~Hermans, M.~Pinzger, and A.~van Deursen.
\newblock Detecting and visualizing inter-worksheet smells in spreadsheets.
\newblock In {\em Proc. ICSE}, pages 441--451, 2012.

\bibitem{6405300}
F.~Hermans, M.~Pinzger, and A.~van Deursen.
\newblock Detecting code smells in spreadsheet formulas.
\newblock In {\em Proc. ICSM}, pages 409--418, 2012.

\bibitem{cunha2012towards}
J{\'a}come Cunha, Jo{\~a}o~P Fernandes, Hugo Ribeiro, and Jo{\~a}o Saraiva.
\newblock Towards a catalog of spreadsheet smells.
\newblock In {\em International Conference on Computational Science and Its
  Applications}, pages 202--216. Springer, 2012.

\bibitem{dou2014spreadsheet}
Wensheng Dou, Shing-Chi Cheung, and Jun Wei.
\newblock Is spreadsheet ambiguity harmful? detecting and repairing spreadsheet
  smells due to ambiguous computation.
\newblock In {\em Proc. ICSE}, pages 848--858. ACM, 2014.

\bibitem{cheung2016custodes}
Shing-Chi Cheung, Wanjun Chen, Yepang Liu, and Chang Xu.
\newblock Custodes: automatic spreadsheet cell clustering and smell detection
  using strong and weak features.
\newblock In {\em Proc. ICSE}, pages 464--475. ACM, 2016.

\bibitem{7884634}
S.~Roy, F.~Hermans, and A.~van Deursen.
\newblock Spreadsheet testing in practice.
\newblock In {\em Proc. SANER}, pages 338--348, 2017.

\bibitem{panko2016we}
Ray Panko.
\newblock What we don't know about spreadsheet errors today: The facts, why we
  don't believe them, and what we need to do.
\newblock {\em arXiv preprint arXiv:1602.02601}, 2016.

\bibitem{rajalingham2008classification}
Kamalasen Rajalingham, David~R Chadwick, and Brian Knight.
\newblock Classification of spreadsheet errors.
\newblock {\em arXiv preprint arXiv:0805.4224}, 2008.

\bibitem{ayalew2008spreadsheet}
Yirsaw Ayalew and Roland Mittermeir.
\newblock Spreadsheet debugging.
\newblock {\em arXiv preprint arXiv:0801.4280}, 2008.

\bibitem{abreu2014faultysheet}
Rui Abreu, J{\'a}come Cunha, Joao~Paulo Fernandes, Pedro Martins, Alexandre
  Perez, and Joao Saraiva.
\newblock Faultysheet detective: When smells meet fault localization.
\newblock In {\em Proc. ICSME}, pages 625--628. IEEE, 2014.

\bibitem{ruthruff2006interactive}
Joseph~R Ruthruff, Margaret Burnett, and Gregg Rothermel.
\newblock Interactive fault localization techniques in a spreadsheet
  environment.
\newblock {\em IEEE Transactions on Software Engineering}, 32(4):213--239,
  2006.

\bibitem{Hermans:2013:DCD:2486788.2486827}
Felienne Hermans, Ben Sedee, Martin Pinzger, and Arie~van Deursen.
\newblock Data clone detection and visualization in spreadsheets.
\newblock In {\em Proc. ICSE}, pages 292--301, 2013.

\bibitem{hermans2013data}
Felienne Hermans, Ben Sedee, Martin Pinzger, and Arie~van Deursen.
\newblock Data clone detection and visualization in spreadsheets.
\newblock In {\em Proc. ICSE}, pages 292--301. IEEE Press, 2013.

\bibitem{roy2009detection}
Chanchal~K Roy.
\newblock Detection and analysis of near-miss software clones.
\newblock In {\em Proc. ICSM}, pages 447--450. IEEE, 2009.

\bibitem{Hermans:2011:SPS:1985793.1985855}
Felienne Hermans, Martin Pinzger, and Arie van Deursen.
\newblock Supporting professional spreadsheet users by generating leveled
  dataflow diagrams.
\newblock In {\em Proc. ICSE}, pages 451--460, 2011.

\bibitem{hermans2010automatically}
Felienne Hermans, Martin Pinzger, and Arie Van~Deursen.
\newblock Automatically extracting class diagrams from spreadsheets.
\newblock {\em Proc. ECOOP}, pages 52--75, 2010.

\bibitem{igarashi1998fluid}
Takeo Igarashi, Jock~D Mackinlay, Bay-Wei Chang, and Polle~T Zellweger.
\newblock Fluid visualization of spreadsheet structures.
\newblock In {\em Proc. ICIV}, pages 118--125. IEEE, 1998.

\bibitem{shiozawa19993d}
Hidekazu Shiozawa, Ken-ichi Okada, and Yutaka Matsushita.
\newblock 3d interactive visualization for inter-cell dependencies of
  spreadsheets.
\newblock In {\em Proc. ICIV}, pages 79--82. IEEE, 1999.

\bibitem{fowler1999refactoring}
Martin Fowler and Kent Beck.
\newblock {\em Refactoring: improving the design of existing code}.
\newblock Addison-Wesley Professional, 1999.

\bibitem{hoepelman2015tool}
DJ~Hoepelman.
\newblock Tool-assisted spreadsheet refactoring and parsing spreadsheet
  formulas.
\newblock 2015.

\end{thebibliography}

\end{document}